\documentclass[useAMS,usedcolumn,usenatbib,usegraphicx]{mn2e}

\usepackage{epsfig,graphicx,natbib}
\usepackage{./reference/mycite}
\usepackage{amssymb}
\usepackage{amsfonts}
\usepackage{amsmath}
\usepackage{color}
 
\begin{document}

\title[2SLAQ LRG Evolution]{Exploring the Luminosity Evolution and Stellar Mass Assembly of 2SLAQ Luminous Red Galaxies Between Redshifts 0.4 and 0.8}
\author[M. Banerji et al.]{ \parbox{\textwidth}{
Manda Banerji$^{1,3}$ \thanks{E-mail: mbanerji@ast.cam.ac.uk},
Ignacio Ferreras$^{2}$, 
Filipe B. Abdalla$^{1}$,
Paul Hewett$^{3}$, 
Ofer Lahav$^{1}$
}
  \vspace*{6pt} \\
$^{1}$Department of Physics and Astronomy, University College London,
Gower Street, London, WC1E 6BT, UK.\\
$^{2}$Mullard Space Science Laboratory, University College London,
Holmbury St Mary, Dorking, Surrey RH5 6NT, UK.\\
$^{3}$Institute of Astronomy, University of Cambridge, Madingley Road,
Cambridge, CB30HA, UK. \\
}

\maketitle

\begin{abstract}

We present an analysis of the evolution of 8625 Luminous Red Galaxies between $z=0.4$ and $z=0.8$ in the 2dF and SDSS LRG and QSO (2SLAQ) survey. The LRGs are split into redshift bins and the evolution of both the luminosity and stellar mass function with redshift is considered and compared to the assumptions of a passive evolution scenario. We draw attention to several sources of systematic error that could bias the evolutionary predictions made in this paper. While the inferred evolution is found to be relatively unaffected by the exact choice of spectral evolution model used to compute K+e corrections, we conclude that photometric errors could be a source of significant bias in colour-selected samples such as this, in particular when using parametric maximum likelihood based estimators. We find that the evolution of the most massive LRGs is consistent with the assumptions of passive evolution and that the stellar mass assembly of the LRGs is largely complete by $z\sim0.8$. Our findings suggest that massive galaxies with stellar masses above $10^{11}$M$_\odot$ must have undergone merging and star formation processes at a very early stage ($z\gtrsim 1$). This supports the emerging picture of \textit{downsizing} in both the star formation as well as the mass assembly of early type galaxies. Given that our spectroscopic sample covers an unprecedentedly large volume and probes the most massive end of the galaxy mass function, we find that these observational results present a significant challenge for many current models of galaxy formation.  

\end{abstract}

\begin{keywords}

galaxies:evolution - galaxies:formation - galaxies:high-redshift - galaxies:luminosity function, mass function

\end{keywords}

\section{Introduction}

Luminous Red Galaxies (LRGs) are arguably some of the brightest galaxies in our Universe allowing us to study their evolution out to much higher redshifts than is possible with samples of more typical galaxies. LRGs are known to form a spectroscopically homogenous population that can be reliably identified photometrically by means of simple colour selections \citep{Eisenstein:01,Cannon:2SLAQ}. The presence of a strong 4000 \AA break in these galaxies also means that accurate photometric redshifts can be derived for large samples where spectroscopy is difficult to obtain \citep{Padmanabhan:05, Collister:MegaZ, Abdalla:MegaZ}. Consequently, samples of luminous red galaxies have emerged as an important data set both for studies of cosmology \citep{Blake:LRG07,Blake:LRG08,Cabre:LRG09} and galaxy formation and evolution \citep{Wake:06,Cool:08}. 

The study of massive galaxies in our Universe is particularly interesting since these galaxies present a long-standing problem for models of galaxy formation. While in the standard $\Lambda$CDM cosmology, massive galaxies are thought to be built up through successive mergers of smaller systems \citep{WhiteRees:78}, many observations now suggest that these galaxies were already well assembled at high redshifts \citep{Glazebrook:04, Cimatti:06, Scarlata:07, Ferreras:09, Pozzetti:09}. At the same time, star formation and merging activity has also been seen in such systems between $z\sim1$ and the present day \citep{Lin:04, Stanford:04} and the studies of \citet{Bell:04} and \citet{Faber:07} have found evidence that the luminosity density of massive red galaxies has remained roughly constant since z$\sim$1 implying a build up in the number density of these objects through mergers. Most observational results now support downsizing in star formation - i.e most massive galaxies having lower specific star formation rates than their less massive counterparts - and many galaxy formation models are able to reproduce these observations e.g. by invoking quenching mechanisms such as AGN feedback \citep{Croton:06, DeLucia:06}. However, the more recently observed trend of downsizing in the mass assembly \citep{Cimatti:06, Pozzetti:09} presents more of a challenge for galaxy formation models which typically predict that the most massive galaxies were assembled later than their less massive counterparts.  

Many of the contradictions in observational studies of massive galaxy formation arise principally for two reasons. Firstly, the way in which early-type galaxies are selected in different surveys can be considerably different. Morphological selection compared to colour selection of such objects will almost certainly result in different samples being chosen, particularly at high redshifts. The colour selection is also usually different for different samples of massive galaxies as we will point out later in this paper. Secondly, many of the studies of massive galaxy formation mentioned so far are deep and narrow spectroscopic samples that will suffer from biases due to the effects of cosmic variance. In addition, one has to be careful in interpreting observational results as in current galaxy formation models, star formation and mass assembly in galaxies are not necessarily concomitant. So while the stars in early-type galaxies may have formed very early in the Universe's history, they may have formed in relatively small units and only merged at lower redshifts to create the massive ETGs we see today. 

Studies of the evolution of luminous red galaxies such as those analysed in this paper have already supported the evidence that massive galaxies have been passively fading and show very little recent star formation \citep{Wake:06, Cool:08}. In this paper, we extend this work by also considering the mass assembly of these systems. The 2dF and SDSS LRG and QSO (2SLAQ) survey presents a comprehensive improvement in volume for massive galaxy samples. The survey covers an area of 186 deg$^2$ and reaches out to a redshift of 0.8 making it a promising data set to study the evolution of massive red galaxies. Furthermore, much of the 2SLAQ area is now covered by the UKIDSS Large Area Survey (LAS) that provides complementary data in the near infra-red bands for the optically selected LRGs. The advantage of near infra-red data is that the mass-to-light ratios and k-corrections are largely insensitive to the galaxy or stellar type and therefore the total infra-red flux in for example the K-band provides a good estimate of the total stellar mass of the galaxies. This stellar mass estimate allows us to study not only the star formation history but also the mass assembly history of these systems. As is the case for optically selected spectroscopic samples of massive galaxies, the K-band selected surveys have so far been restricted to relatively small and deep patches of the sky \citep{Mignoli:K20, Conselice:07}. Clearly a large spectroscopic survey of massive galaxies with optical and near infra-red photometry, will allow for better constraints on the evolution of these systems. 

In this paper, we utilise a spectroscopic sample of colour-selected massive red galaxies from the 2SLAQ survey between redshifts 0.4 and 0.8. Optical photometry is obtained from the Sloan Digital Sky Survey supplemented with near infra-red data from the UKIDSS Large Area Survey (LAS). We consider the evolution of these galaxies in terms of their observed colours, luminosities and comoving number densities focussing particularly on the effects of colour selection on the inferred evolution.  \citet{Wake:06} have presented a comprehensive analysis of the luminosity function of Luminous Red Galaxies using data from both the Sloan Digital Sky Survey and the 2SLAQ Survey. These authors use a subset of data from the 2SLAQ survey in the redshift range 0.5 to 0.6 and by comparing this galaxy population to a lower redshift population at $0.17<z<0.24$ from SDSS, they are able to establish that the LRG LF does not evolve beyond that expected from a simple passive evolution model at these redshifts. Meanwhile, \citet{Cool:08} have compared the low-redshift SDSS LRG population to their own high-redshift sample at redshifts of $\sim 0.9$ and once again find little evidence for evolution beyond the passive fading of the stellar populations. We extend the work of these authors by considering now most of the galaxies available in the 2SLAQ data set. This enables the redshift range of 2SLAQ galaxies used to be broadened to $0.4\leq z < 0.8$.  The 2SLAQ luminosity function is presented for 8625 LRGs as compared to the 1725 used by \citet{Wake:06}. These results are useful in filling the gap between redshift 0.4 and 0.5 and redshift 0.6 and 0.8 in the \citet{Wake:06} and \citet{Cool:08} data sets. The sample is split into four redshift bins and the evolution of the luminosity and colour of LRGs between these redshift bins is considered. In addition, we also consider the stellar mass function and its evolution with redshift. Furthermore, we use various different stellar population synthesis models that are commonly used in the literature, to model the LRGs, including the new models of \citet{Maraston:09} and quantify the sensitivity of the luminosity function estimate to changes in these models. The optical colours of LRGs have long been difficult to model using standard spectral evolution models \citep{Eisenstein:01} and this problem has only recently been solved \citep{Maraston:09} thereby allowing us to utilise the most accurate spectral evolution models of LRGs to date in order to infer their evolution. 

The paper is structured as follows. In $\S$ \ref{sec:data} we describe the spectroscopic 2SLAQ data set as well as SDSS and UKIDSS photometry for these galaxies. $\S$ \ref{sec:models} describes the spectral evolution models used in this paper to model the LRGs. We consider the optical luminosity function of the 2SLAQ LRGs in $\S$ \ref{sec:lf} and its evolution with redshift as well as its sensitivity to cosmic variance, photometric errors and changes in the spectral evolution models. In $\S$ \ref{sec:number} we present estimates for the LRG mass function in different redshift bins and analyse its sensitivity to the choice of spectral evolution model as well as the IMF. Finally, we discuss our results in $\S$ \ref{sec:disc} in terms of models of galaxy formation and evolution. Throughout this paper we assume a cosmological model with $\Omega_m$=0.3, $\Omega_\Lambda$=0.7 and $h$=0.7. All magnitudes are in the AB system unless otherwise stated. 

\section{DATA}

\label{sec:data}

Our dataset is a spectroscopic sample of Luminous Red Galaxies from the 2dF and SDSS LRG and QSO (2SLAQ) survey \citep{Cannon:2SLAQ}. This survey was conducted on the 2-degree Field (2dF) spectrograph on the 3.9m Anglo-Australian Telescope. The survey recorded spectra for $\sim$10000 LRGs with a median redshift of 0.55 and $\sim$10000 faint $z<3$ QSOs selected from SDSS imaging data. The survey covers an area of 186 deg$^2$ and extends to a redshift of $\sim0.8$ for the LRGs. In this section, we provide a description of the 2SLAQ data as well as optical and near infra-red photometry for these galaxies obtained using the SDSS and UKIDSS LAS \citep{Smith:09} respectively. 

In the 2SLAQ survey, the following colour and magnitude cuts have been applied using SDSS DR4 photometry in order to isolate the main LRG population to target with spectroscopy \citep{Cannon:2SLAQ}.

\begin{equation}
c_{\parallel}=0.7(g-r)+1.2(r-i-0.18) \ge 1.6
\label{eq:cut3}
\end{equation}

\begin{equation}
d_{\perp}=(r-i)-(g-r)/8.0>0.65
\label{eq:cut4}
\end{equation}

\begin{equation}
17.5 \leq i_{deV} - A_i \leq 19.8
\label{eq:cut5}
\end{equation}

\noindent where $g$, $r$ and $i$ denote the dereddened SDSS model magnitudes, $i_{deV}$ is the i-band deVaucouleurs magnitude and $A_i$ is the extinction in the i-band. The deVaucouleurs magnitudes are obtained by fitting a pure deVaucouleurs profile to the 2D galaxy images. The definitions of both these types of magnitudes as well as other SDSS parameters can be found at \texttt{http://www.sdss.org/dr4/algorithms/photometry.html} as well as \citet{Stoughton:02}.  

These cuts are for the primary sample defined as Sample 8 in \citet{Cannon:2SLAQ} which has the highest completeness and this is the sample that we use throughout this paper. M stars are expected to make up $\sim$5\% of the sample but can be excluded by their low redshifts. In addition, as will be discussed later in the paper, photometric errors could scatter objects both in and out of the sample across the colour selection boundaries and lead to some contamination. \citet{Roseboom:06} have conducted a detailed study of the star formation histories of this 2SLAQ LRG sample and find that 80\% of the objects, which represents the vast majority, are passive in nature as would be expected from early type systems. The rest are mainly emission line galaxies. Note also that the colour cuts applied in order to isolate LRGs are usually different for different surveys. For example the colour cuts used to isolate LRGs in the SDSS LRG sample \citep{Eisenstein:01} results in a distribution that is redder than that of the 2SLAQ galaxies. If these different LRG samples are to be compared therefore, further cuts must be applied to them to ensure a consistent colour selection as in \citet{Wake:06}. 

In the case of the 2SLAQ LRGs, the effect of the $d_{\perp}$ cut is to select early-type galaxies at increasingly high redshifts whereas the $c_\parallel$ cut eliminates late-type galaxies from the sample. Most of the sample has redshifts of $0.2<z<0.8$ with $\sim$5\% contamination from M-type stars \citep{Cannon:2SLAQ}. In order to capture the main redshift distribution of this primary sample, only galaxies with $0.4 \leq z<0.8$ are used in this work. Note, however that most of the galaxies lie above a redshift of 0.45 and our conclusions remain unchanged if we choose the lower redshift limit of our sample to be 0.45 instead. Furthermore, we have also selected only galaxies with a redshift quality flag of greater than 2.  This results in a sample of 8625 LRGs and the redshift completeness for this sample is 76.1\% (David Wake:private communication). All LRGs in our sample have secure spectroscopic redshifts. 

\begin{figure*}
\begin{center}
\begin{minipage}[c]{1.00\textwidth}
\centering
\includegraphics[width=8.5cm,angle=0]{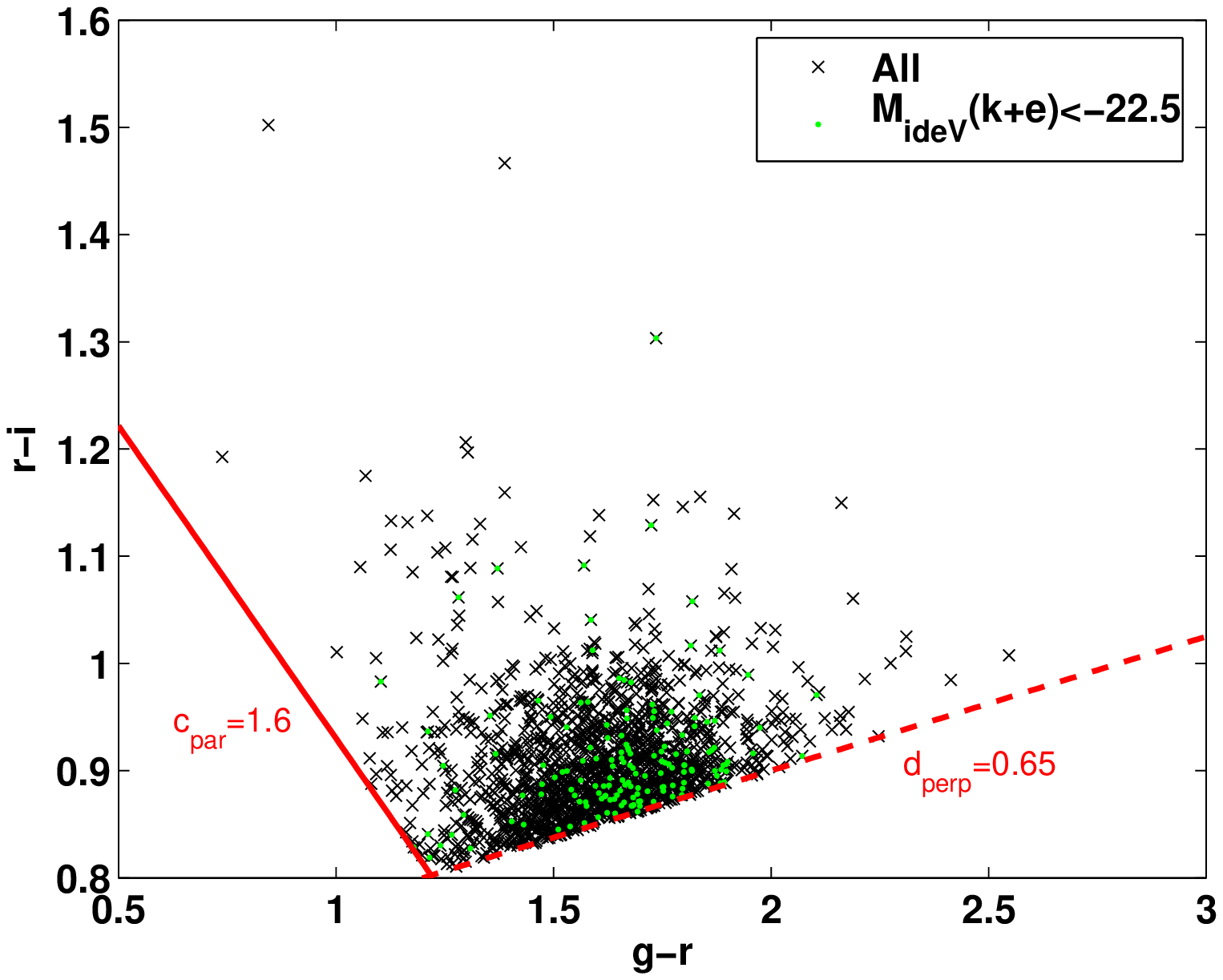}
\includegraphics[width=8.5cm,angle=0]{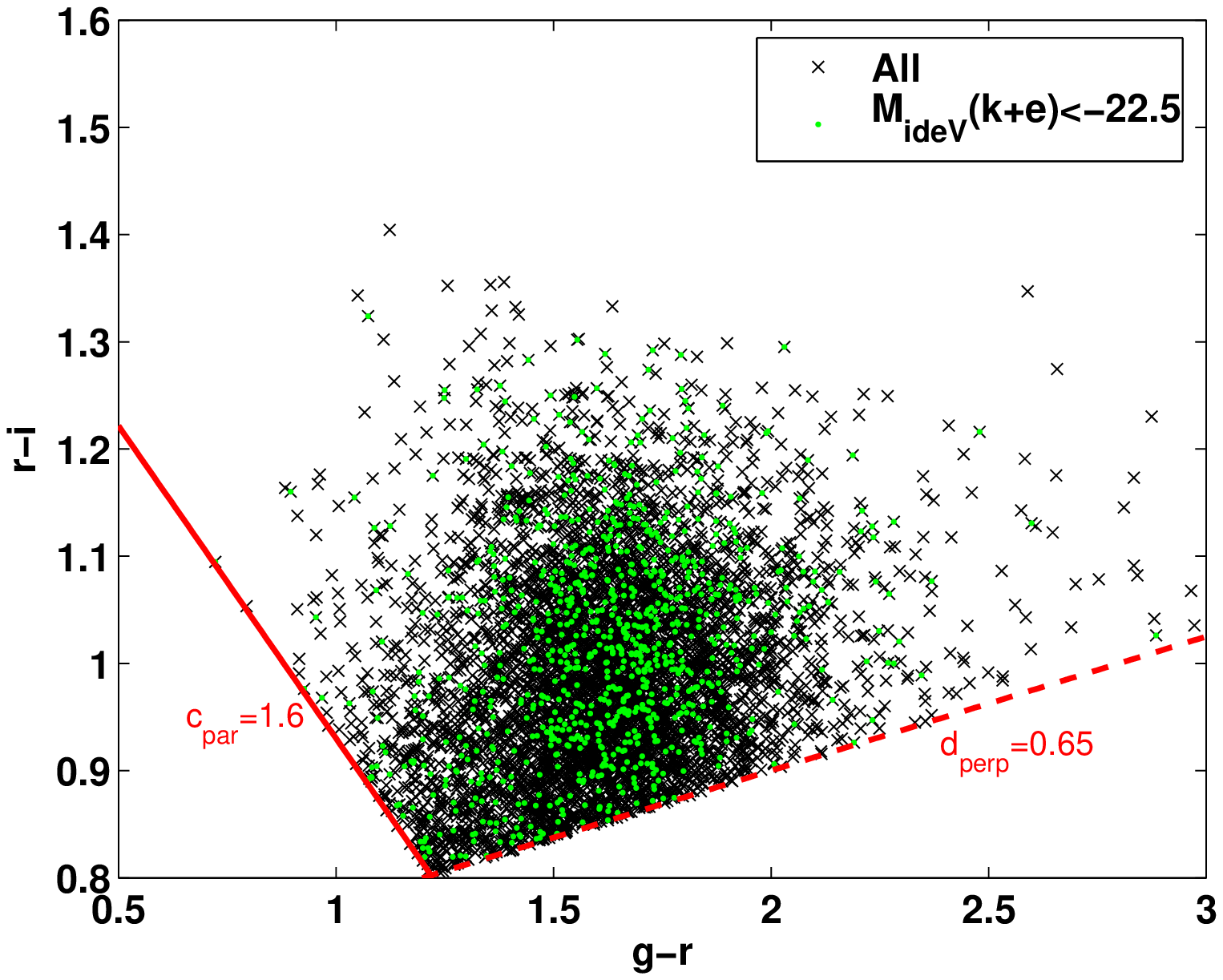}
\includegraphics[width=8.5cm,angle=0]{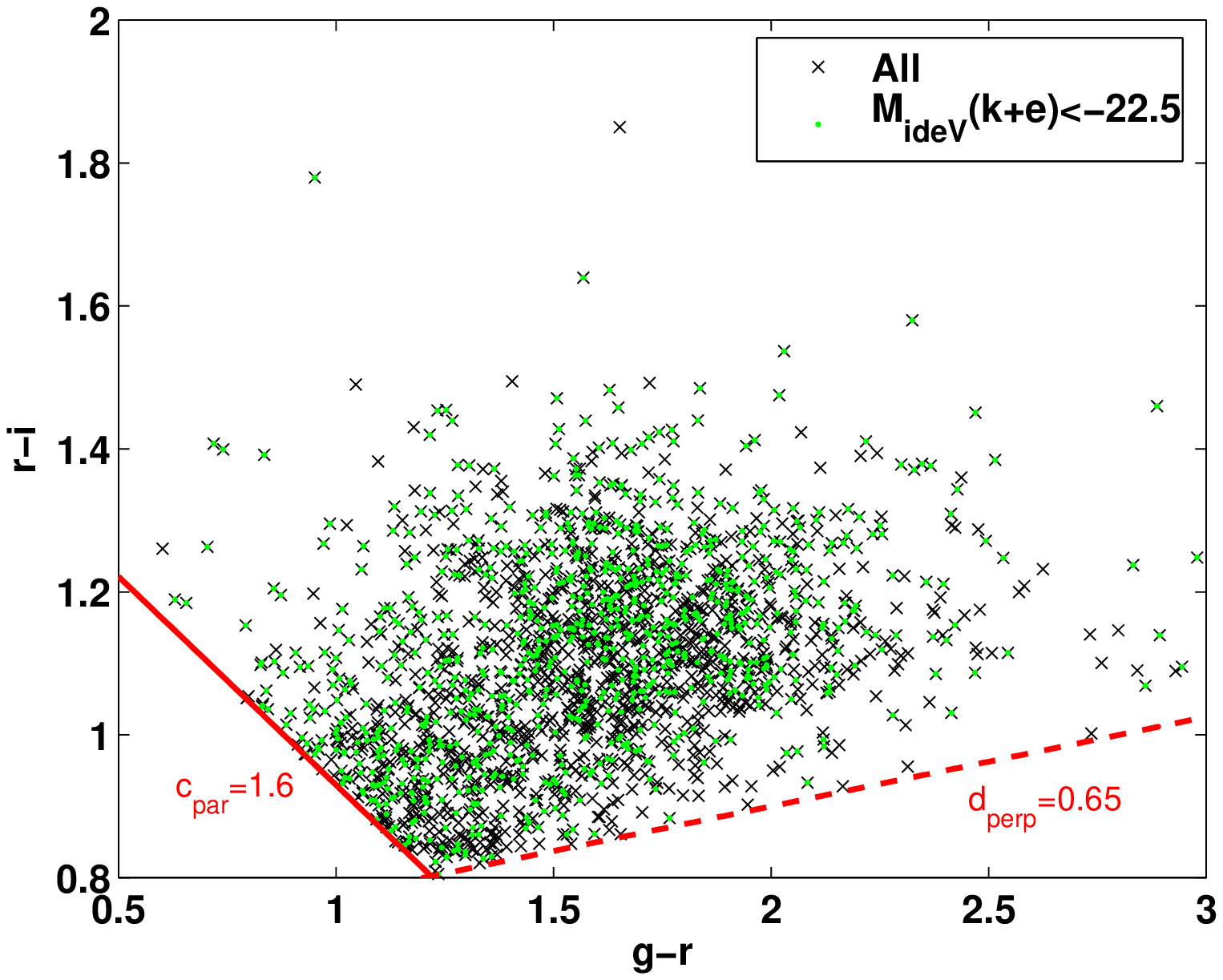}
\includegraphics[width=8.5cm,angle=0]{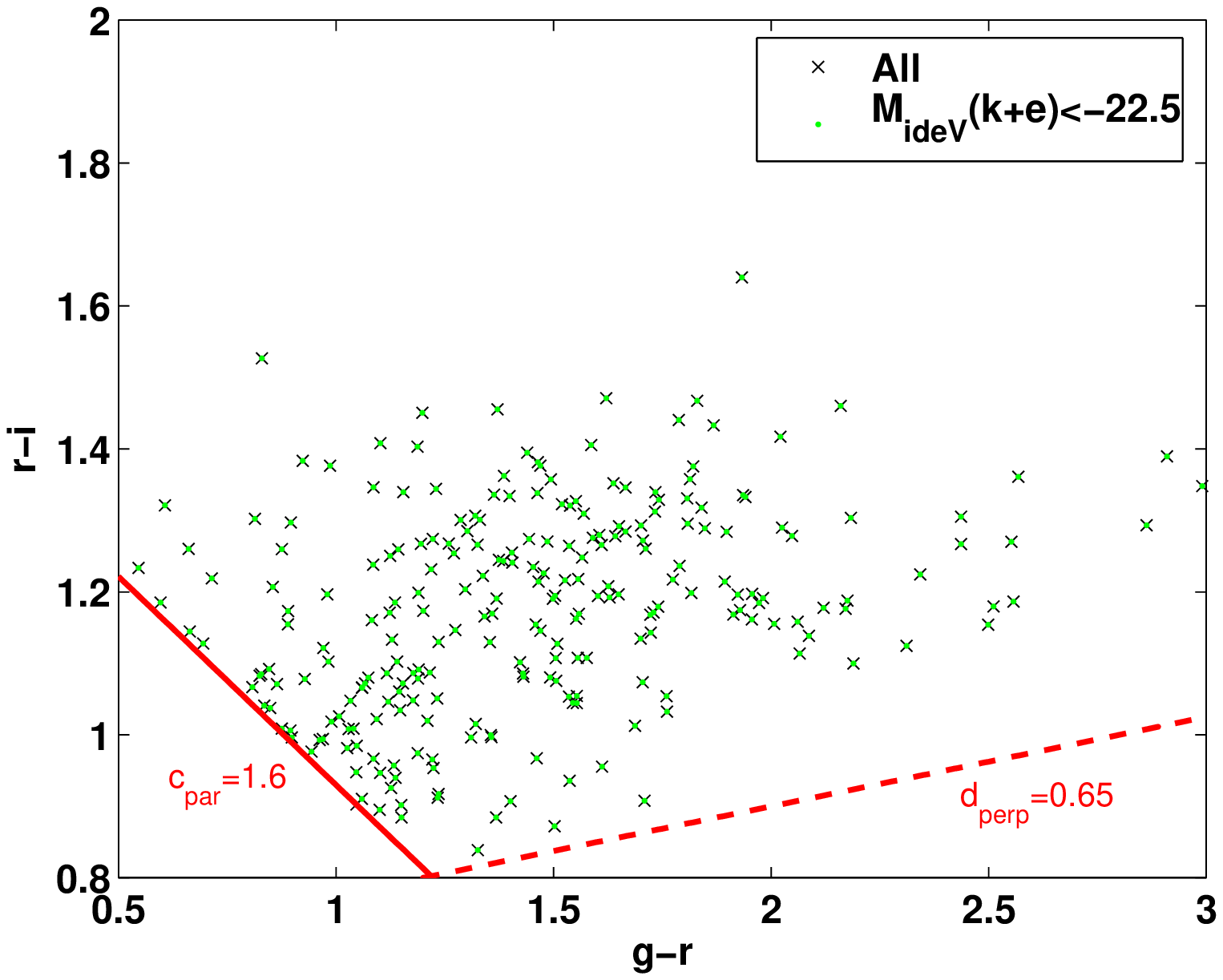}
\end{minipage}
\caption{The (g-r) versus (r-i) colours of 2SLAQ galaxies in the observed frame in four redshift bins - $0.4 \leq z < 0.5$ (top left), $0.5 \leq z < 0.6$ (top right), $0.6 \leq z < 0.7$ (bottom left) and $0.7 \leq z < 0.8$ (bottom right). The dark crosses correspond to all galaxies in that redshift bin whereas the light dots show the brightest galaxies with $M_{ideV}<-22.5$.}
\label{fig:colourz}
\end{center}
\end{figure*} 

In Figure \ref{fig:colourz}, we plot the LRGs in the observed (g-r) versus (r-i) plane in redshift bins of width 0.1. Isolating bright galaxies only with $M_{ideV}<-22.5$ and plotting them in the same colour-colour plane shows clearly how the red sequence is truncated in the lowest redshift bin due to the redshift dependant colour selection. The effect of this truncation on the luminosity function estimate will be described in more detail in $\S$ \ref{sec:lf}. 

In order to obtain accurate stellar masses for the LRGs, the 2SLAQ data is matched to near infra-red data from the UKIDSS Large Area Survey (LAS) DR5 which has a depth of $K_{Vega}$=18.4. The galaxies are matched to within 1.1 arcsecs in position using the WFCAM science archive\footnote{http://surveys.roe.ac.uk/wsa/index.html} and only galaxies with detections in the K-band are selected. Out of our total sample of 8625 2SLAQ LRGs, we have K-band data for 6476 of them which represents $\sim$75\% of the sample. The UKIDSS K-band magnitudes are in the Vega system and have been corrected to the AB system using the corrections of \citet{Hewett:06}. 

\subsection{Photometric Errors}

\label{sec:photerr}

\citet{Wake:06} have conducted a detailed analysis of the effect of photometric errors on the 2SLAQ LRG sample. These photometric errors could induce a systematic bias in the sample by scattering galaxies both in and out of the sample across the colour selection boundaries. For $i<19.3$, the effect of the photometric errors is relatively insignificant and the colour-magnitude distributions of LRGs selected using single-epoch photometry and those selected using multi-epoch photometry are almost identical \citep{Wake:06}. Fainter than this however, the photometric errors may present a significant systematic bias to the inferred colour and luminosity evolution of LRGs in this survey. Note that the magnitude errors inferred using the single-epoch data are found to be systematically underestimated. The effect of these photometric errors on the luminosity function estimate is considered in $\S$ \ref{sec:err}. It turns out that different estimators of the luminosity function have different sensitivity to the photometric errors and this point is addressed in more detail in Appendix A. 

\section{K+e Corrections}

\label{sec:models}

In order to calculate the rest-frame properties of LRGs and infer their evolution, the observed properties need to be transformed into the rest-frame by means of a k-correction. In addition, one can also correct for any evolutionary changes expected in the galaxy spectra by means of an e-correction. Motivated by the work of \citet{Padmanabhan:05}, we start by using a Pegase \citep{Fioc:PEGASE} template to model the spectral evolution of the LRGs. In this template, the stars are formed in a single burst 11Gyr ago ($z\simeq2.5$). We assume a \citet{Salpeter:IMF} IMF, solar metallicity and no galactic winds or substellar objects. 

However, it has been noted by several authors that stellar population synthesis models such as Pegase and those of \citet{BC:03} fail to reproduce the observed colours of LRGs \citep{Eisenstein:01, Maraston:05}. Improvements to these models have subsequently been made that involve changes to the input stellar libraries as well as including an additional metal-poor sub component to the stellar population in order to create a composite model \citep{Maraston:09}. Such models predict significantly bluer (g-r) colours compared to simple stellar population models and are better able to reproduce the observed broadband colours of individual LRGs. For this reason, a Maraston model of age 12Gyr corresponding to a galaxy mass of 10$^{12}$ M$_\odot$ is also considered in this work. 

Finally, we consider the models of Charlot \& Bruzual 2007 \citep{BC:03,Bruzual:07} - CB07 hereafter - with a range of different metallicities and star formation histories in order to quantify the systematics associated to the different prescriptions, isochrones and stellar libraries of the population synthesis models. All the CB07 templates assume a formation redshift of z$_{\rm F}=3$ and a \citet{Chabrier:03} IMF. 

\begin{figure*}
\begin{center}
\begin{minipage}[c]{1.00\textwidth}
\centering
\includegraphics[width=8.5cm,angle=0]{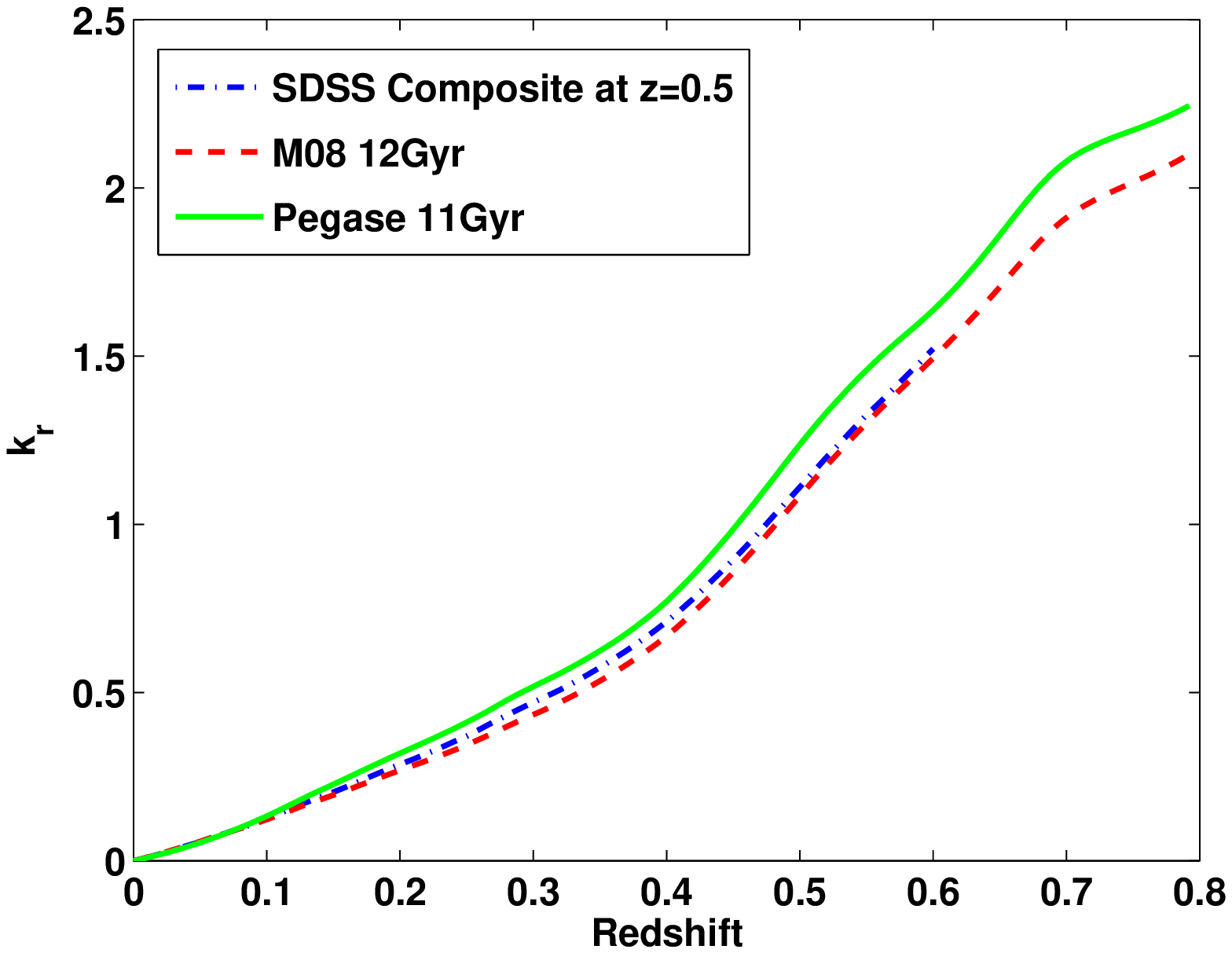}
\includegraphics[width=8.5cm,angle=0]{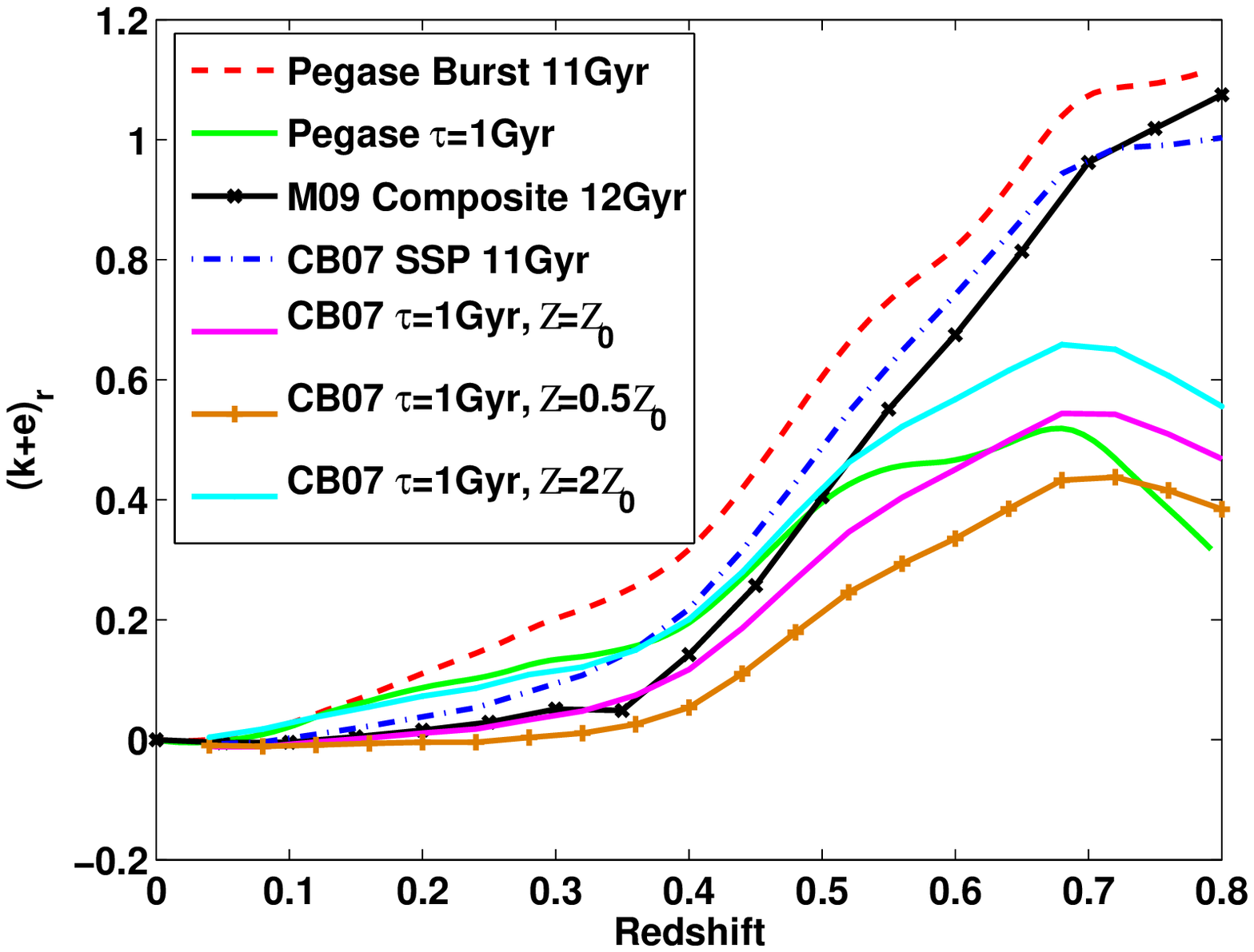}
\end{minipage}
\caption{The k (left) and k+e (right) corrections in the $r$-band for LRGs derived using stacked SDSS spectra as well as various spectral evolution models.}
\label{fig:kcorr}
\end{center}
\end{figure*}

K and k+e corrections in the r-band derived from each of these models, are illustrated in Figure \ref{fig:kcorr}. In addition, k-corrections are also derived from stacked LRG spectra from the SDSS LRG survey \citep{Eisenstein:01} at redshifts 0.2 and 0.5. These are very similar to each other suggesting that there is little or no colour evolution in the SDSS LRGs between redshifts of 0.2 and 0.5. As the 2SLAQ LRG spectra are not flux calibrated, a similar derivation of the k-corrections using these was not possible. Although the SDSS LRGs are selected using different colour cuts, it is still interesting to compare their k-corrections to the ones derived using the various spectral evolution models and it can be seen that the empirical k-corrections best match those derived from a Maraston model. For this reason, the composite 12Gyr model from \citet{Maraston:09} has been used throughout this paper unless otherwise stated. We can also see from the right-hand panel of Figure \ref{fig:kcorr} that the e-correction is very sensitive both to the star-formation history of the galaxy as well as the metallicity as galaxies with some residual star formation as well as those at high metallicity have more negative e-corrections. 

\begin{figure*}
\begin{center}
\begin{minipage}[c]{1.00\textwidth}
\centering
\includegraphics[width=8.5cm,angle=0]{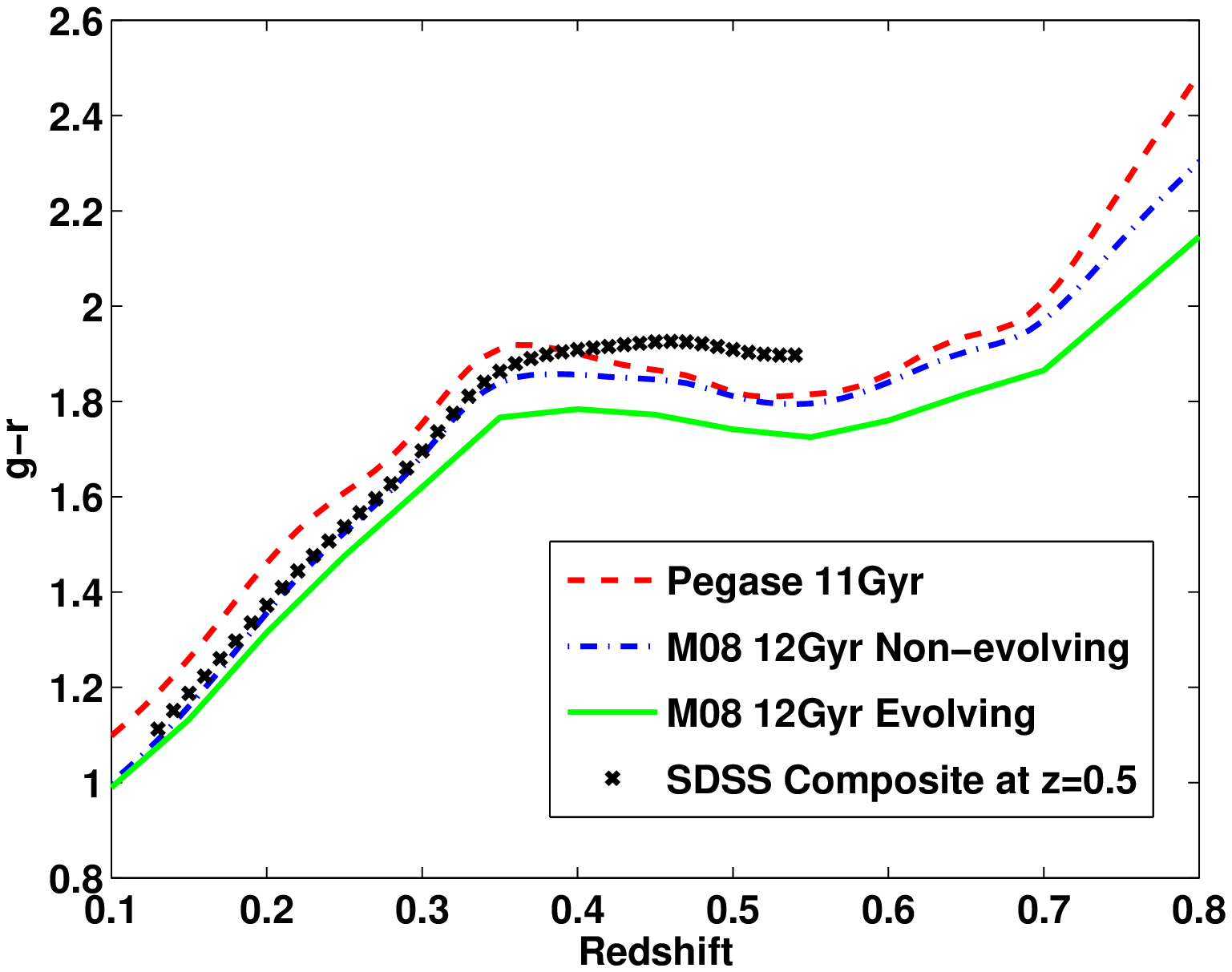}
\includegraphics[width=8.5cm,angle=0]{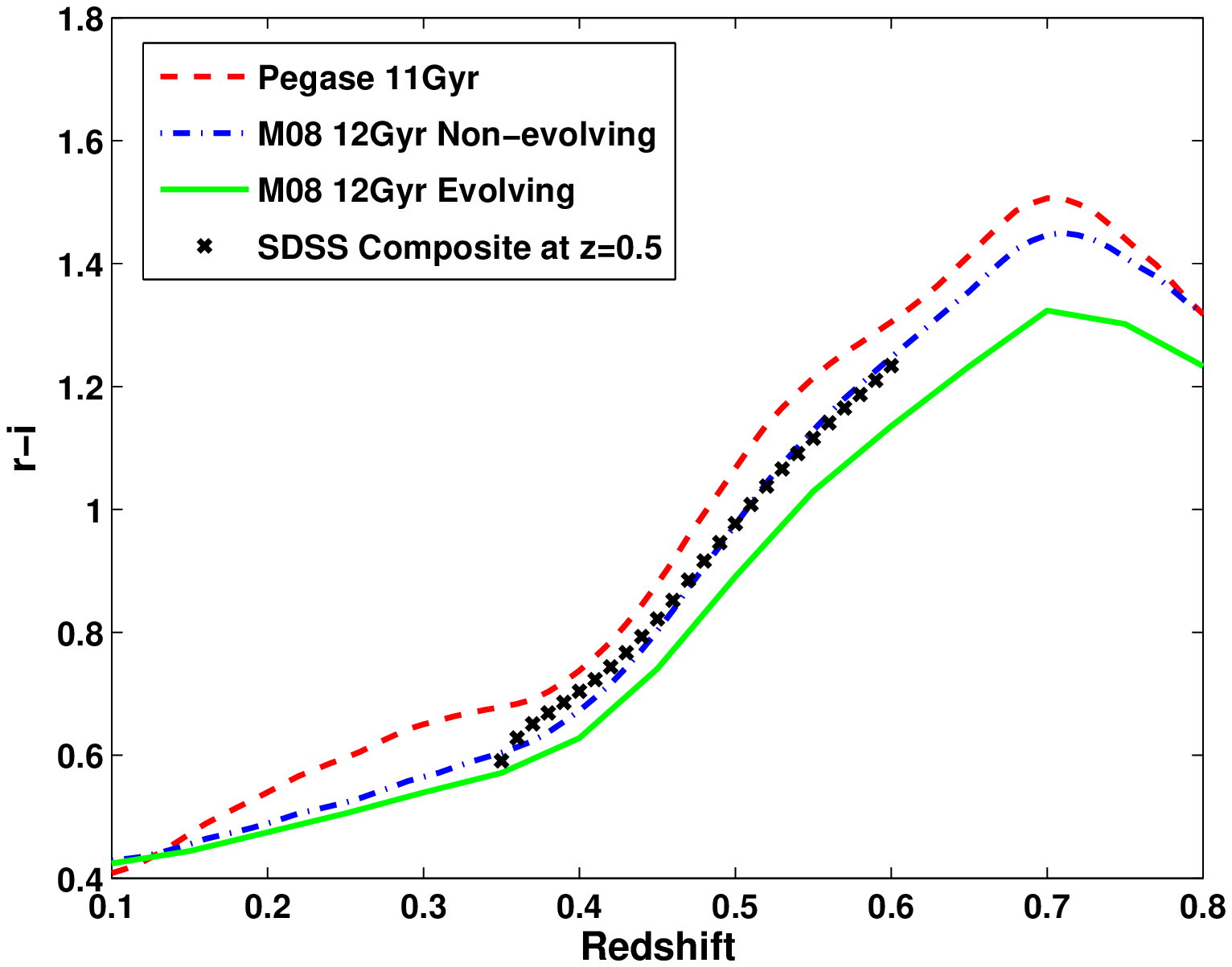}
\end{minipage}
\caption{The (g-r) - left - and (r-i) -right - colours of LRGs as a function of redshift.}
\label{fig:colours}
\end{center}
\end{figure*}

In Figure \ref{fig:colours} we also plot the rest-frame (g-r) and (r-i) colours of LRGs derived using the different spectral evolution models as a function of redshift. Once again it is seen that the Maraston model is able to reproduce the colours derived from SDSS composite spectra almost exactly unlike the Pegase model. However, we note that unlike in \citet{Maraston:09}, we find that including the evolutionary correction to the Maraston models, creates a mismatch in the colours of the model and the SDSS composite spectra. The SDSS observed colours are found to be redder than the evolving Maraston model. We have already noted that the SDSS LRG colour distribution is redder than that of the 2SLAQ galaxies used in \citet{Maraston:09} for comparison to their models so this difference is perhaps to be expected. It should also be noted that the observed colours derived from the spectra are derived from the fiber magnitudes of the galaxies. As the fibers only enclose light within a 3'' aperture and it is well known that early type galaxies have colour gradients \citep{Vader:88, Franx:90, Ferreras:05}, this could be why the colours appear redder but it is unlikely that this effect would be significant especially at redshifts of 0.5. Finally, the redshift 0.5 LRGs in the SDSS survey are the brightest and therefore expected to be the reddest galaxies in this survey \citep{Bernardi:05, Gallazzi:06}. All these factors may help explain the discrepancy in the observed and model colours derived using the evolving Maraston model. 

\begin{figure*}
\begin{center}
\includegraphics[width=12cm,angle=0]{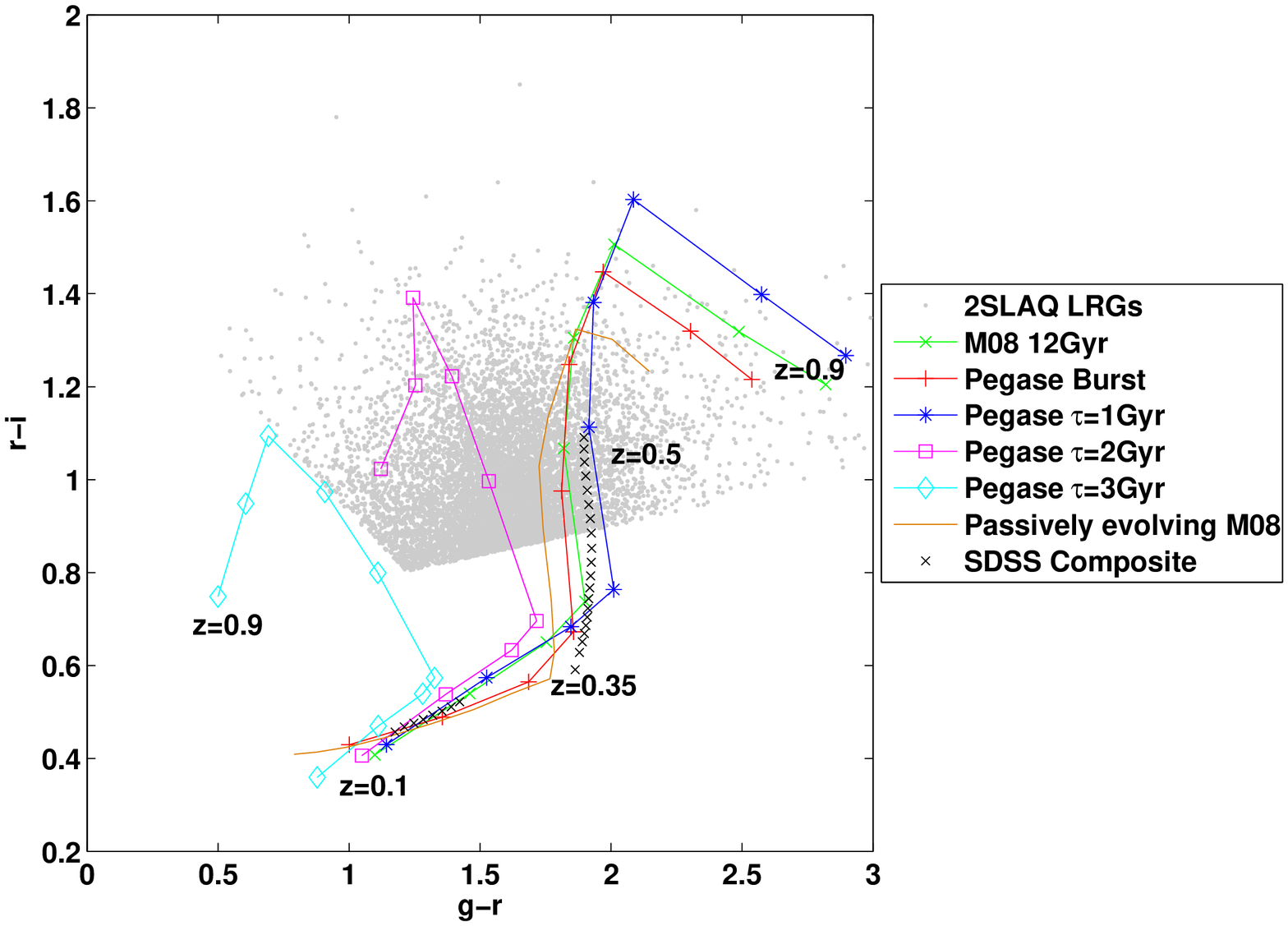}
\caption{The (g-r) versus (r-i) colours of 2SLAQ galaxies in the observed frame along with tracks produced by several different stellar population synthesis models as well as composite LRG spectra from SDSS. In the case of the models, the markers are located at redshift intervals of 0.1. In the case of the SDSS composite spectrum, the markers are located at redshift intervals of 0.01.}
\label{fig:gri}
\end{center}
\end{figure*}

Figure \ref{fig:gri} shows the 2SLAQ LRGs in the observed (g-r) versus (r-i) colour-colour plane along with the main colour selection boundaries as well as the tracks produced in this plane by different stellar population synthesis models as well as SDSS composite spectra. It can be seen that both the Maraston as well as Pegase burst models only enter the $d_\perp$ selection at redshifts of $\sim$0.45. Star forming galaxies with $\tau>2$Gyr are clearly excluded at all redshifts by the $c_\parallel$ selection.

\section{THE OPTICAL LUMINOSITY FUNCTION}

\label{sec:lf}

The luminosity function for 2SLAQ LRGs is calculated in this section in order to infer any evolution in the observed number density of these objects. The non-parametric $V_{max}$ estimator is used to calculate luminosity functions and a gaussian parametric form is fit by means of a $\chi^2$ minimisation where appropriate. The gaussian parametric form provides a better fit to the luminosity function of early-type galaxies compared to the more commonly used Schechter function and has three free parameters, $\phi_*$, $M_*$ and $\sigma_g$. The parametric form is given by:

\begin{equation}
\Phi(M)=\frac{\phi_*}{\sigma_g \sqrt{2\pi}}\exp \left(\frac{-(M-M_*)^2}{2\sigma_g^2}\right).
\label{eq:gauss}
\end{equation} 

The commonly used parametric estimator of \citet{STY:79} (STY hereafter) is prone to biases for this data set and therefore has not been presented in the main body of the paper. Details of the STY estimator and the biases induced on it are given in Appendix A. 

The V$_{max}$ estimator relies on measuring the maximum volume occupied by a galaxy given the survey selection criteria. By scaling the observed number density by this accessible volume, we account for the fact that only the brightest galaxies are observed at high redshifts in a flux-limited sample. For the 2SLAQ survey, this maximum volume depends not only on the maximum redshift out to which the galaxy could be observed given the flux limit of $i_{dev} - A_i < 19.8$ but also the minimum redshift at which the galaxy would be included in the survey given the $d_\perp$ and $c_\parallel$ colour selection criteria. The maximum volume for each galaxy is then given by: 

\begin{equation}
V_{max} = \int_{0}^{\infty} S_{d_\perp}(z) S_{c_\parallel}(z) S_{i_{deV}}(z) \frac{dV}{dz} dz
\label{eq:vmax}
\end{equation}

\noindent where $S_{d_\perp}(z)$, $S_{c_\parallel}(z)$ and  $S_{i_{deV}}(z)$ are the selection functions for each galaxy due to the two primary colour cuts (Eq \ref{eq:cut3} and \ref{eq:cut4}) and the flux limit (Eq \ref{eq:cut5}). Once this maximum volume has been derived for every galaxy in the sample, the luminosity function which is simply the number density of galaxies per unit brightness, is given by: 

\begin{equation}
\Phi(M) dM = c \sum_{i}{\frac{1}{V_{max,i}}}
\label{eq:phi_vmax}
\end{equation}

\noindent where c is the redshift completeness of the sample assumed to be 76.1\% and the index, $i$ runs over all galaxies in the absolute magnitude bin $dM$.

The errors on the luminosity function are assumed to be Poissonian and are given by:

\begin{equation}
\sigma_{\Phi}(M) dM = c \sqrt{\sum_{i}{\frac{1}{V_{max,i}^{2}}}}
\label{eq:error_vmax}
\end{equation}

The best-fit parametric form is derived using a $\chi^2$ fit to the $V_{max}$ data points. We marginalise over the normalisation, $\phi_*$ which is determined independantly from the observed number density of galaxies. The marginalised $\chi^2$ to be minimised is:

\begin{equation}
\chi^2_{marg}=\frac{\sum_j \sum_i \sigma_{V,i}^2 \Phi_{m,j} \Phi_{V,j} (\Phi_{m,i} \Phi_{V,j} - \Phi_{m,j} \Phi_{V,i}) / \sigma_{V,j}^2}
                   {\sum_j \Phi_{m,j}^2 / \sigma_{V,j}^2}
\label{eq:chi2}
\end{equation}

\noindent where $\Phi_{m}$ refers to the parametric luminosity function which is assumed to be a gaussian function and $\Phi_V$ represents the $V_{max}$ data points with errors $\sigma_V$. Once the best fit parameters have been determined,  the normalisation of the luminosity function, $\phi_*$ is then determined by matching the parametric estimate of the luminosity function to the observed number density of galaxies using Eq. \ref{eq:norm}. 

\begin{equation}
\phi_*=\frac{N_g}{\int dz f_s \frac{dV}{dz} \int dM f(M,z) \Phi(M)}
\label{eq:norm}
\end{equation} 

\noindent where $N_{g}$ is the number of galaxies, $f_s$ is the fraction of sky covered by the survey and $f(M,z)$ represents the colour selection function obtained by summing the product of $S_{d_\perp}(z)$ and $S_{c_\parallel}(z)$ over all galaxies in the sample. The errors on the gaussian parameters $M_*$ and $\sigma$ are calculated using the 1$\sigma$ error ellipsoid obtained by performing the minimisation using a Markov Chain Monte Carlo (MCMC) method. The main sources of error on $\phi_*$ come from cosmic variance and the covariance of $\phi_*$ with $M_*$. In $\S$ \ref{sec:cosmicvar} we will show that for a large survey such as 2SLAQ, cosmic variance is unlikely to be a dominant systematic. The errors for $\phi_*$ quoted throughout the paper are therefore calculated from the 1$\sigma$ errors on $M_*$. 

The dependance of the luminosity function on redshift is first considered and possible sources of systematic errors to these estimates examined later in this section. 

\subsection{Redshift Evolution}

\label{sec:redshift}

We examine the redshift evolution of the 2SLAQ LRG luminosity function in four redshift bins between redshift 0.4 and 0.8. The luminosity function estimates are obtained using the $V_{max}$ estimator described above. Due to the different absolute magnitude ranges in each redshift bin, no $\chi^2$ fits to the data points are shown as the inferred parameters would not be comparable. 

\begin{figure*}
\begin{center}
\begin{minipage}[c]{1.00\textwidth}
\centering
\includegraphics[width=8.5cm,angle=0]{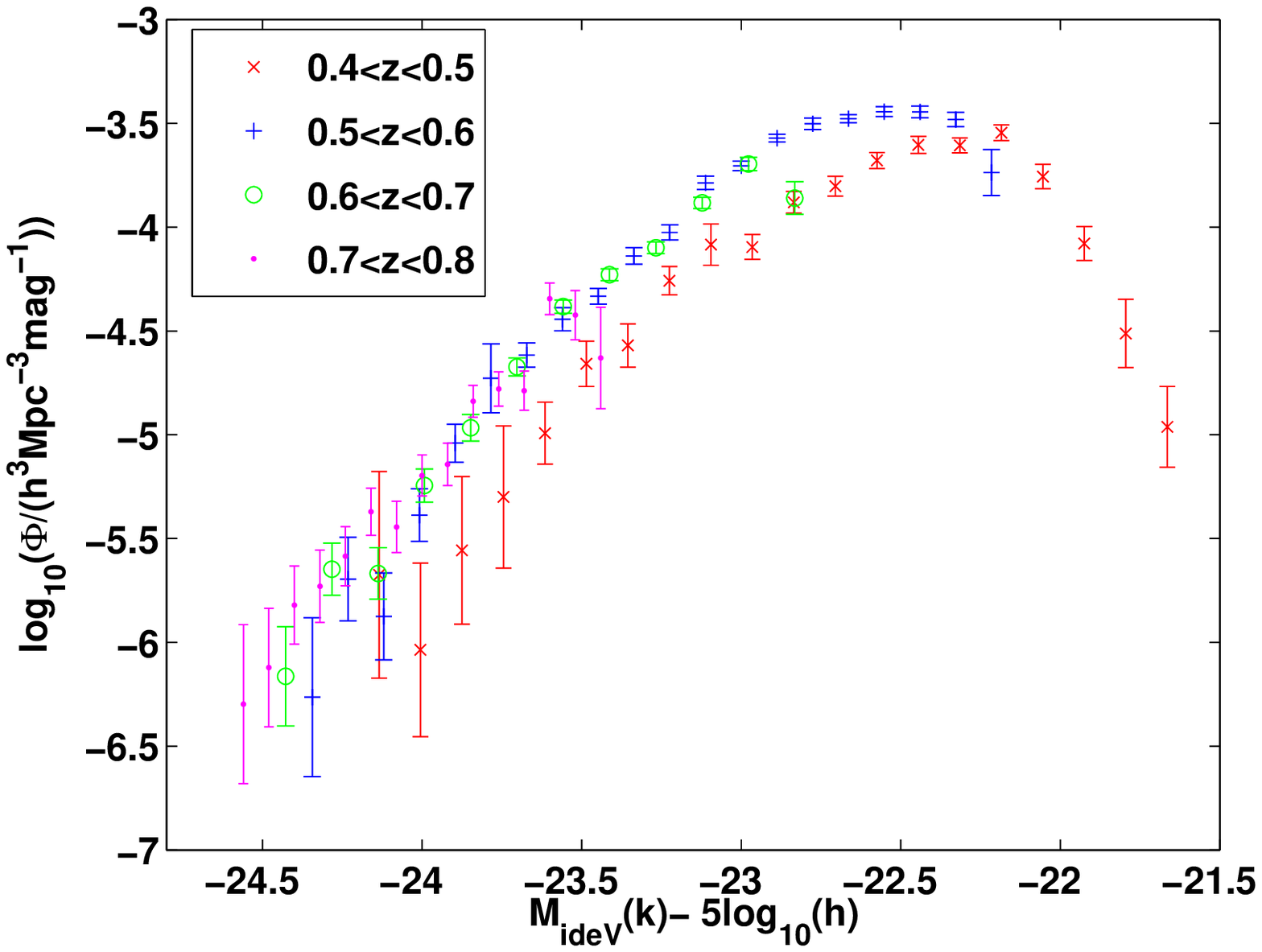}
\includegraphics[width=8.5cm,angle=0]{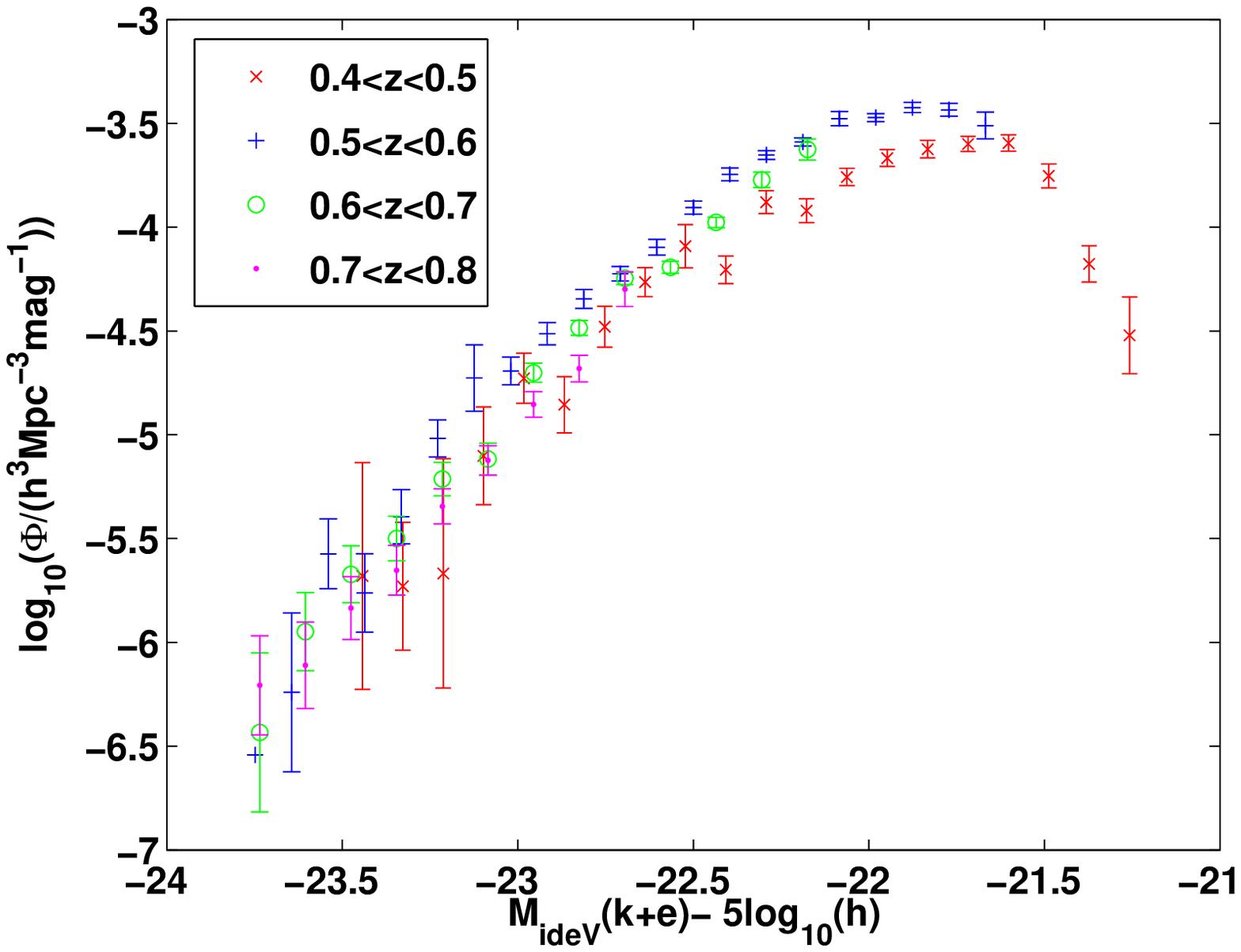}
\end{minipage}
\caption{The 2SLAQ luminosity function at a rest-frame redshift 0 after k-correction (left) and k+e correction (right) assuming passive evolution.}
\label{fig:lfbins}
\end{center}
\end{figure*}

\begin{figure}
\begin{center}
\includegraphics[width=8.5cm,angle=0]{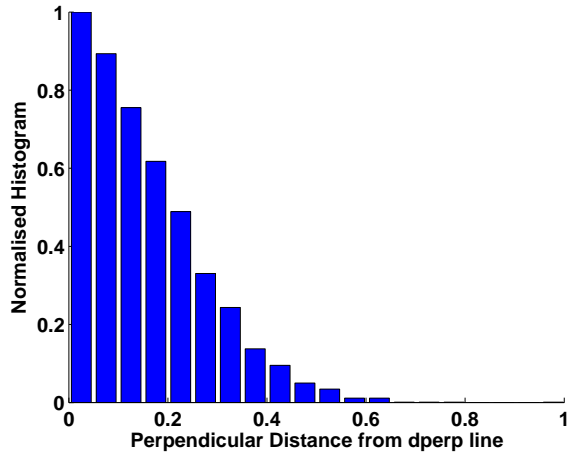}
\caption{Normalised histograms of the perpendicular distance of 2SLAQ galaxies with $i_{deV}>19$ from the d$_\perp$ selection line.}
\label{fig:distance}
\end{center}
\end{figure}

In the left-hand panel of Figure \ref{fig:lfbins} we plot the luminosity function in four redshift bins after K correcting to redshift 0 using the 12Gyr Maraston composite model. The number density of galaxies is found to increase slightly with redshift particularly in the brightest absolute magnitude bins. In the faintest bins there is a drop in the observed number density. As LRGs are known to be some of the brightest objects in the Universe and our survey performs an LRG selection, we would expect a drop in the number density of such systems at faint magnitudes. Downsizing in the star formation of early type systems, is an effect that is already well observed. The fainter galaxies are expected to be less massive and therefore bluer due to their higher specific star formation rates and such objects are excluded from our sample by the colour selection criteria which only select out the brightest and reddest objects. The V$_{max}$ method used to evaluate luminosity functions already accounts for those galaxies that may be missing from our sample due to the observational limits set by the survey. It does not however account for photometric uncertainties in the sample which we study in detail in $\S$ \ref{sec:err}. It is possible that genuine LRGs may be scattered out across the colour selection boundaries due to their large photometric errors rendering the sample incomplete at the faint end. In order to demonstrate that the downturn we see at the faint-end of our luminosity function estimates is not due to any such incompleteness, we plot in Figure \ref{fig:distance} the normalised histogram of the perpendicular distance of the faintest 2SLAQ LRGs with $i_{deV}>19$ from the d$_\perp$ selection line. Most of these objects that appear faint and are likely to have large photometric errors, are seen to lie very close to the selection line - i.e. at a distance of $\sim$0 . This distribution therefore suggests that there are also likely to be large numbers of faint galaxies on the other side of the d$_\perp$ selection. Any photometric errors would thus scatter more galaxies into the sample than are being scattered out resulting in an overprediction in the number density of the faint galaxies that are close to the flux limit of the survey. We therefore conclude that the faint-end downturn seen in our luminosity function estimates is genuine and not due to any incompleteness induced by the photometric errors. We have also seen in Figure \ref{fig:colourz} how the red sequence is truncated in the lowest redshift bin due to the redshift dependant colour selection. This is what leads to a drop in the number density in the lowest redshift bin relative to the other bins which is not due to any evolution in the underlying galaxy population. 

In the right-hand panel of Figure \ref{fig:lfbins}, we plot the luminosity functions after correcting for passive evolution assuming the same Maraston model. The luminosity functions change little on inclusion of this evolutionary correction suggesting it is small in these redshift ranges but in general the data points for galaxies with $M_{ideV} < -22.5$ in the different redshift bins now agree better. We are confident that incompleteness does not affect this bright galaxy sample and these results therefore suggest that this LRG population is evolving consistently with the assumptions of a passive evolution model.  

\subsection{Sources of Systematic Error}

In this section we examine potential sources of systematic error that could bias the luminosity function estimates presented above.

\subsubsection{Photometric Errors}

\label{sec:err}

In order to assess the bias that photometric errors will introduce into the luminosity function estimate, we cut our sample at $i<19.3$ and calculate the luminosity function for this reduced sample of galaxies. Note that in $\S$ \ref{sec:photerr} we have already drawn attention to the fact that previous studies \citep{Wake:06} have shown that fainter than $i=19.3$, galaxy samples selected through single and multi-epoch photometry differ considerably. The single-epoch magnitude errors are also found to be systematically underestimated. 

\begin{figure}
\begin{center}
\includegraphics[width=8.5cm,angle=0]{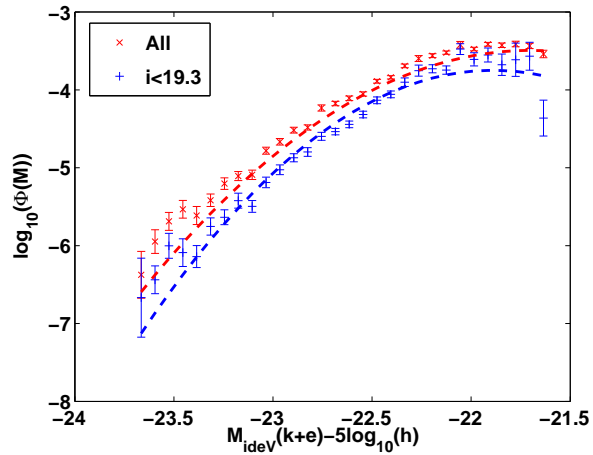}
\caption{The 2SLAQ luminosity function calculated for the entire sample and galaxies with $i<19.3$ for which photometric errors are thought to be insignificant along with gaussian fits to both. Only the bright-end of the luminosity function for the entire sample is shown in order to ensure that the gaussian fits are made over the same absolute magnitude range.}
\label{fig:photerr}
\end{center}
\end{figure}

The luminosity functions for the entire sample as well as the $i<19.3$ sample are plotted in Figure \ref{fig:photerr} along with gaussian fits to these derived using $\chi^2$ minimisation. Only the bright-end of the luminosity function for the entire sample is shown as the sample with small photometric errors has no galaxies in the faintest absolute magnitude bins and we want to compare the two luminosity functions over the same absolute magnitude range. The $i<19.3$ sample has 2131 galaxies as opposed to the 8625 galaxies in the entire sample. The space densities from the $i<19.3$ sample have therefore been renormalised by the fraction of galaxies in the complete sample that contribute at each absolute magnitude bin. The $\chi^2$ fit to the reduced sample with small photometric errors yields values of $\sigma_g=0.45 \pm 0.02$, $M_*=-21.89 \pm 0.05$ and $\phi_*=2.64 \pm ^{0.21} _{0.18} \times 10^{-4}$ where the normalisation, $\phi_*$ for the $i<19.3$ sample has also been corrected for the smaller space density of this sample. In this case however, a constant average correction factor is used rather than the absolute magnitude dependent corrections applied to the $V_{max}$ data points in order to ensure that the shape of the gaussian is maintained. The $\chi^2$ fit to the entire sample on the other hand gives $\sigma_g=0.52 \pm 0.01$, $M_*=-21.70 \pm 0.04$ and $\phi_*=5.51 \pm ^{0.53} _{0.47} \times 10^{-4}$. 

As can be seen, the faintest galaxies in terms of absolute magnitude are removed from our sample on applying the $i$-band cut, as they typically have larger photometric errors. However, the two luminosity functions are remarkabely similar in shape and there is evidence for a faint-end downturn even for the sample with small photometric errors. The space density at the bright-end is also lower for our sample with small photometric errors even after renormalisation suggesting that photometric errors affect galaxies at all absolute magnitudes. The space density for the entire sample is found to be about a factor of two larger due to galaxies being scattered into the sample due to their large photometric errors. We have already shown that more galaxies are likely to be scattered in across the colour selection boundaries than are scattered out (Figure \ref{fig:distance}). These galaxies are bluer than more typical LRGs but appear red due to the photometric errors. Many of their spectra show evidence for the presence of strong emission lines suggesting that they are star-forming and this has already been found by \citet{Roseboom:06}. 

\subsubsection{Cosmic Variance}

\label{sec:cosmicvar}

\begin{figure}
\begin{center}
\includegraphics[width=8.5cm,angle=0]{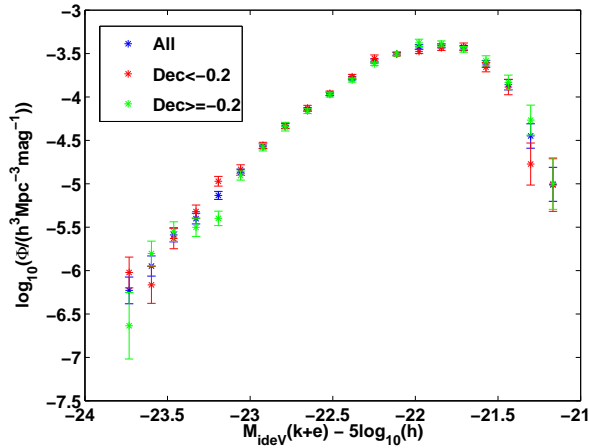}
\caption{The 2SLAQ luminosity function calculated for the entire sample and in two different patches in the sky in order to assess the effects of cosmic variance.}
\label{fig:cosmicvar}
\end{center}
\end{figure}

One of the main drawbacks of the V$_{max}$ estimator is that it can suffer from biases due to cosmic variance. As massive galaxies such as the ones in this sample are thought to be strongly clustered, the evolution in their number density will be very sensitive to large-scale structure. Traditionally, this has posed a problem for many small surveys but the 2SLAQ survey should cover a significant enough volume to make this an unlikely source of problem for the LRGs being studied. In order to test this hypothesis, we split the LRG sample by declination and seperate galaxies with dec $<-0.2$ and dec $\geq -0.2$. This results in two smaller samples with 4340 and 4285 galaxies respectively. A luminosity function is then calculated for each of these subsamples as well as the total luminosity function for the 2SLAQ sample. These are plotted in Figure \ref{fig:cosmicvar}. No notable differences can be seen in the three luminosity function estimates confirming that cosmic variance is not a dominant systematic in this sample. 

\subsubsection{K+e Corrections: Spectral Evolution Model}

\begin{figure*}
\begin{center}
\includegraphics[width=20cm,height=12cm,angle=0]{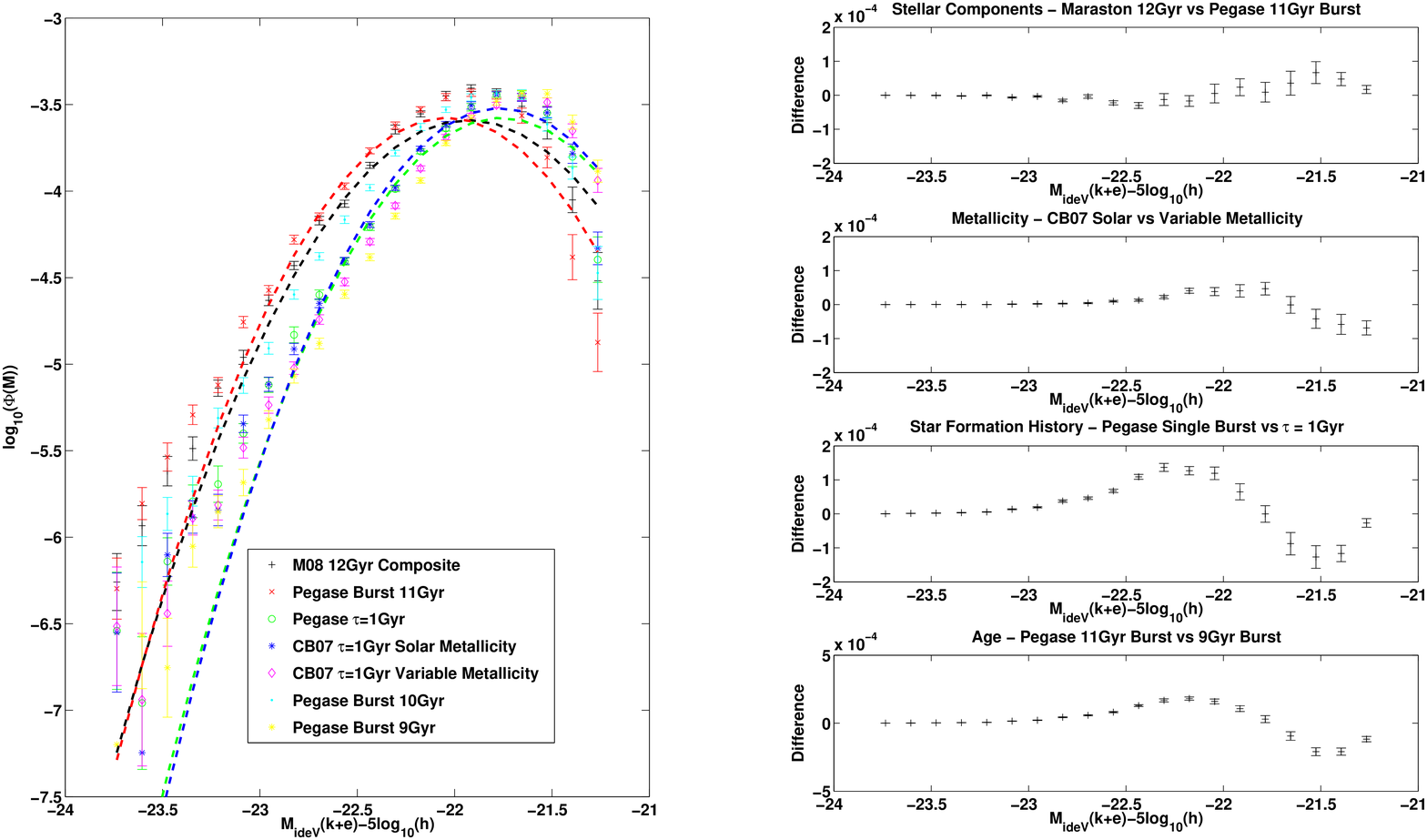}
\caption{The 2SLAQ luminosity function computed using K+e corrections from different spectral evolution models with different prescriptions, isochrones and stellar libraries (left). Gaussian fits are also shown to the luminosity functions estimated using the Maraston 12Gyr model, the Pegase 11Gyr burst model, a Pegase model with $\tau$=1Gyr as well as a CB07 model at solar metallicity with $\tau$=1Gyr. The best-fit gaussian parameters are summarised in Table \ref{tab:model}. The right panel shows the difference between the luminosity function estimates as a function of the absolute magnitude when different parameters associated with the spectral evolution models are varied. }
\label{fig:model}
\end{center}
\end{figure*}

In order to test whether the choice of spectral evolution model used to compute K+e corrections significantly affects the estimate of the luminosity function, we compute luminosity functions for all the 2SLAQ galaxies between redshift 0.4 and 0.8 using the 12Gyr Maraston composite model as well as a Pegase burst model, a Pegase model with an exponentially declining star formation history with $\tau=1$Gyr and a CB07 model with an exponentially declining star formation history with $\tau=1$Gyr. Both Pegase models have an age of 11Gyr while the CB07 model has a formation redshift of 3 corresponding to an age 11.35Gyr in our chosen cosmology. The results are illustrated in Figure \ref{fig:model} and the best-fit gaussian parameters to these luminosity functions summarised in Table \ref{tab:model}. Both the Pegase burst and Maraston models have similar ages and star formation histories. The difference between them is the introduction of a metallicity poor sub-component to the Maraston model and the use of improved stellar librarires in order to better match the observed colours of LRGs. However, it can be seen from Figure \ref{fig:model} (left panel and right,top) that there is very little difference in the luminosity function estimate obtained from using these two models. 

\begin{table*}
  \begin{center}	
    \begin{tabular}{c|c|c|c}
      & $\sigma_g$ & $M_*$ & $\Phi_*$ \\
      \hline
      \hline
      Maraston 12Gyr & 0.44 $\pm$ 0.01 & -21.93 $\pm$ 0.03 & 3.73 $\pm ^{0.24} _{0.23}$ $\times$10$^{-4}$ \\
      Pegase Burst 11Gyr & 0.41 $\pm$ 0.01 & -22.04 $\pm$ 0.03 & 3.59 $\pm ^{0.22} _{0.20}$ $\times$10$^{-4}$ \\
      Pegase $\tau$=1Gyr & 0.41 $\pm$ 0.01 & -21.76 $\pm$ 0.03 & 3.58 $\pm ^{0.25} _{0.23}$ $\times$10$^{-4}$ \\ 
      CB07 $\tau$=1Gyr & 0.41 $\pm$ 0.01 & -21.77 $\pm$ 0.03 & 3.98 $\pm ^{0.28} _{0.25}$ $\times$10$^{-4}$\\
      \hline
\end{tabular}	\vspace{2mm}
  \end{center}
  \caption{Summary of best-fit gaussian parameters to the luminosity function derived using different spectral evolution models \label{tab:model}}
\end{table*} 

The Pegase and CB07 models with $\tau$=1Gyr produce very similar luminosity functions despite having different initial mass functions as the K+e corrections and therefore the luminosity function estimate is not very sensitive to the number of low mass stars. The main difference between the Salpeter IMF assumed in the Pegase models and the Chabrier IMF assumed in the CB07 models, is the mass distribution for stars below 1M$_\odot$. Changing the star formation history to have a little residual star formation shifts the luminosity function to slightly fainter absolute magnitudes and the number density in the brightest bins decreases. This is because including recent star formation or younger stars in the model means that the luminosity of the galaxy evolves faster and the galaxy fades quicker than in a passive evolution scenario. 

\subsubsection{K+e Corrections: Metallicity}

The colour-magnitude relation of early-type galaxies can be interpreted as a metallicity sequence and variations in metallicity dominate the slope of the CMR \citep{Kodama:97, Bower:92}. The metallicity may therefore be an important parameter to vary in stellar population synthesis models when trying to model the spectral evolution of LRGs. In order to assess the importance of metallicity in the luminosity function estimate, we calculate luminosity functions after K+e correcting galaxies using a CB07 model at solar metallicity as well as one with variable metallicity. Both assume a formation redshift of 3 and an exponentially declining star formation history with a timescale, $\tau$ of 1Gyr. In the variable metallicity model, the metallicities are derived from the absolute magnitudes using the colour-magnitude relation of early-type galaxies in the Virgo Cluster from \citet{Bower:92} 

\begin{equation}
\log_{10}\left(\frac{Z}{Z_\odot}\right)=-0.11(M_i+21.0)-0.18
\label{eq:virgo}
\end{equation}

The effect of metallicity on the luminosity function estimate is also illustrated in Figure \ref{fig:model}. The second plot in the right panel illustrates the difference between the luminosity function estimates obtained using a CB07 model at solar metallicity and one with variable metallicity. The difference between the two estimates is found to be roughly of the same order as the errorbars on the individual luminosity function estimates. It is therefore concluded that the metallicity is not likely to be important in the evaluation of the luminosity function for the 2SLAQ LRGs studied here. The LRGs are all found to have metallicities between 0.6 and 1.5 $Z_\odot$ with a mean metallicity of roughly solar.  

\subsubsection{K+e Corrections: Age}

In Figure \ref{fig:model} we also plot the V$_{max}$ estimate of the luminosity function after K+e corrections to redshift 0 assuming a Pegase single burst model at three different ages ranging from 9 to 11Gyr. In the cosmological model assumed throughout this paper, these correspond to formation redshifts for the galaxy of between $\sim$ 1.4 and 2.6. We find as expected that changing the formation redshift of the galaxy to later times results in the luminosity function moving to fainter absolute magnitudes or alternatively, a decrease in the number density at a given absolute magnitude. As is the case for models with some residual star formation, the introduction of younger stars into the models means that they fade quicker thereby shifting the luminosity function estimate to fainter absolute magnitudes. 

\section{THE INFRA-RED LUMINOSITY FUNCTION AND STELLAR MASS FUNCTION}

\label{sec:number}

So far we have considered the evolution of the 2SLAQ LRG optical luminosity function with redshift and quantified various sources of systematic error that could arise for this estimate. This analysis shows little evidence that the LRG population evolves beyond that expected from a simple passive evolution model between a redshift of 0.4 and 0.8. However, such an analysis does not fully exploit all the available information as the evolution of elliptical galaxy populations is known to be a strong function of mass \citep{DeLucia:06, Cimatti:06, Ferreras:09, Pozzetti:09}. The availability of near infra-red data for a sizeable subsample of these LRGs means we can calculate accurate stellar masses for these objects from their K-band luminosities in order to see if the evolution of the number density of these galaxies is a strong function of the mass.

The mass function analysis in this section is carried out for the subset of LRGs for which near infra-red photometry is available from UKIDSS DR5. This reduced sample contains 6476 galaxies. As the Maraston models do not extend to near infra-red wavelengths, the CB07 models have been used in this section to calculate the mass-to-light ratios for the LRGs. Following \citet{Ferreras:09}, the stellar masses are computed by comparing the SDSS $g, r, i, z$ fiber magnitudes with a set of 10 $\tau$-models with solar metallicity, formation redshifts of 3 and exponential timescales between 0 (i.e. corresponding to a Simple Stellar Population) and 10Gyr. The CB07 models are used to generate the composite populations. The best-fit model, almost always corresponding to short formation timescales, is then used to determine the mass-to-light ratio in the $K$-band of the UKIDSS survey assuming a Chabrier or Salpeter IMF. This best-fit along with the $K$-band petrosian magnitude allows us to determine the stellar mass of the galaxy. Note that no total model magnitudes are available in the UKIDSS survey. However, the difference between the total model magnitudes in the SDSS $i$-band and the petrosian magnitudes in the same band are found to be of the order of 15\%. Both magnitudes can be used as reasonable estimates for the total light from the galaxy. 

\begin{figure}
\begin{center}
\includegraphics[width=8.5cm,angle=0]{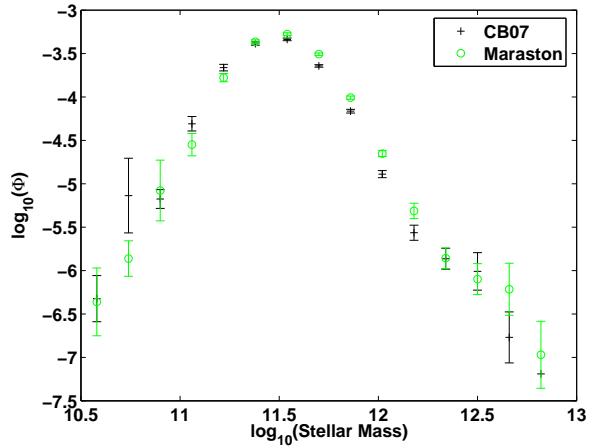}
\caption{The 2SLAQ mass function calculated using the V$_{max}$ method when V$_{max}$ is evaluated using a single Maraston model of age 12Gyr for all the LRGs and when V$_{max}$ is evaluated using the best-fit CB07 $\tau$ model for each LRG.}
\label{fig:mf_model}
\end{center}
\end{figure}

The mass functions are calculated as before using the $V_{max}$ method described in detail in $\S$ \ref{sec:lf}. The difference in the mass function when V$_{max}$ is determined using the best-fit CB07 models as opposed to a single Maraston model of age 12Gyr, is illustrated in Figure \ref{fig:mf_model}. From this figure, it can be seen that the difference is minimal when using the two different spectral evolution models. Note, however that this may not remain the case if stellar masses could be calculated directly from the Maraston models as well as determining $V_{max}$ from it. 

\begin{figure*}
\begin{center}
\begin{minipage}[c]{1.00\textwidth}
\centering
\includegraphics[width=8.5cm,angle=0]{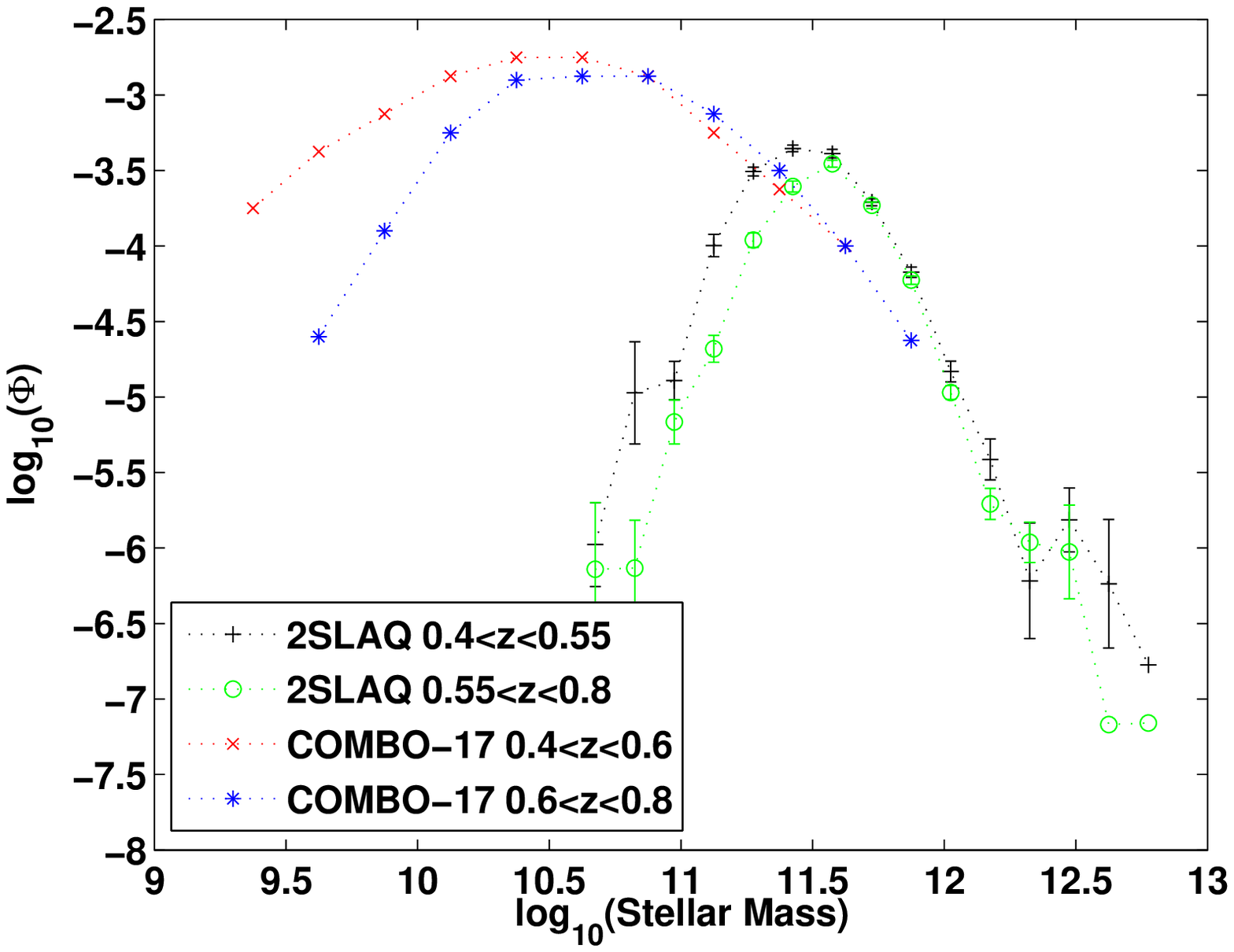}
\includegraphics[width=8.5cm,angle=0]{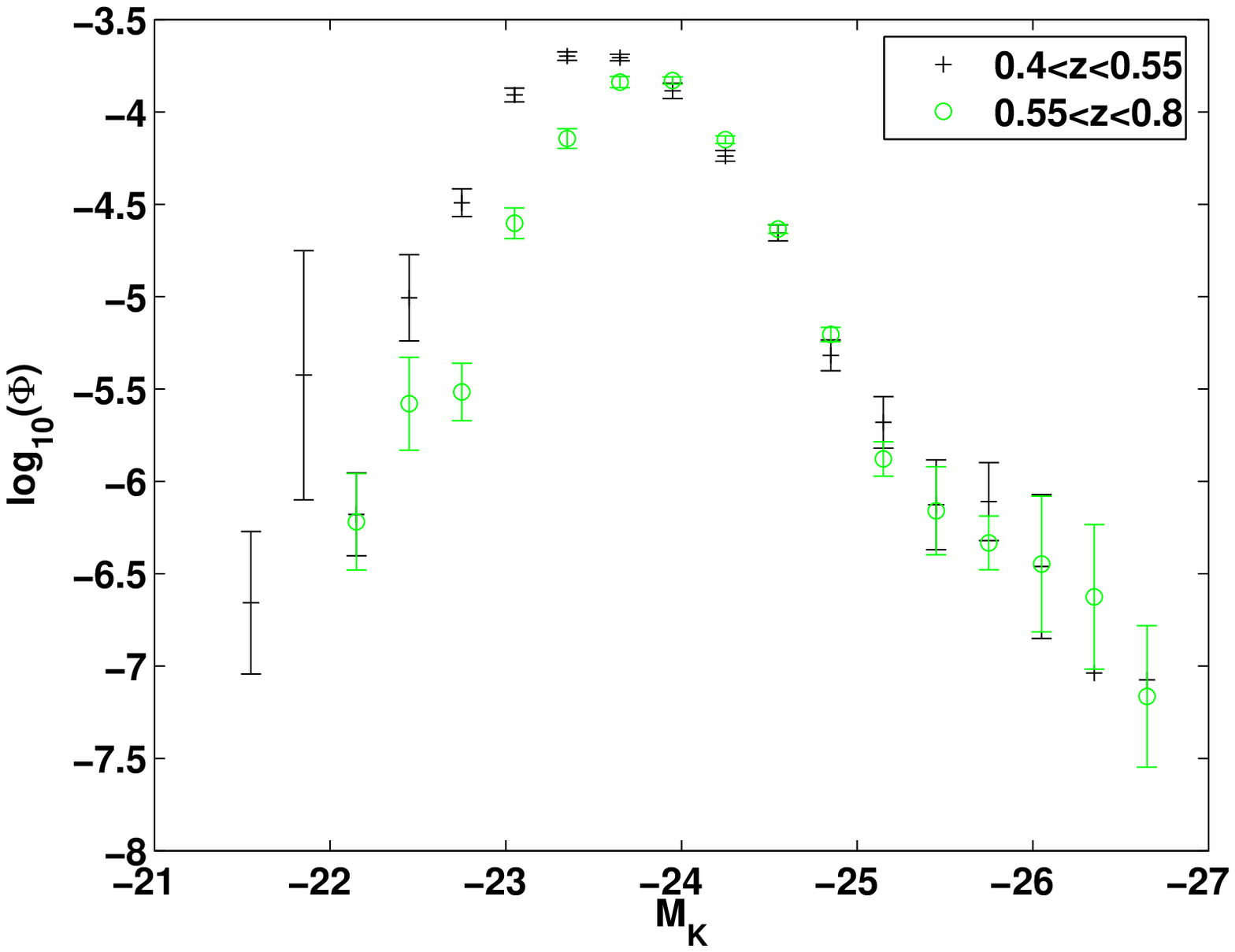}
\end{minipage}
\caption{The 2SLAQ stellar mass function in two redshift bins of - $0.4 \leq z < 0.55$ and $0.55 \leq z < 0.8$ along with the COMBO-17 red sequence stellar mass functions in redshift bins $0.4 \leq z \le 0.6$ and $0.6 \leq z \leq 0.8$ from \citet{Borch:06}. The right-hand panel shows the 2SLAQ K-band luminosity function in the same two redshift bins.}
\label{fig:mf}
\end{center}
\end{figure*}

The mass function for the 2SLAQ LRGs in two redshift bins - $0.4 \leq z \le 0.55$ and $0.55 \leq z \le 0.8$ - is illustrated in the left-hand panel of Figure \ref{fig:mf}. These redshift limits have been chosen so as to ensure that incompleteness does not affect the high-mass end in either of these bins. On this plot, we also show the mass functions derived in two redshift bins for red sequence galaxies in the COMBO-17 survey (from Figure 9 of \citet{Borch:06}). For illustrative purposes, we plot the K-band luminosity function in the two redshift bins in the right-hand panel of Figure \ref{fig:mf}. The K-band absolute magnitudes are also determined using the best-fit CB07 model for each LRG and after correcting to the AB system using the correction of \citet{Hewett:06}. 

Figure \ref{fig:mf} clearly shows that the 2SLAQ data set extends to much higher stellar masses than the COMBO-17 survey occupying a very different stellar mass range to the COMBO-17 red sequence and thereby allowing us to effectively probe the high-mass end of the mass function. These most massive galaxies are the ones expected to place the most stringent constraints on current models of galaxy formation. Our sample also probes a volume that is $\sim$200 times bigger than that of COMBO-17 over the redshift range $0.4 \leq z < 0.8$. As is the case with the COMBO-17 data, there is little evidence for any evolution in the number density at the high-mass end between a redshift of 0.4 and 0.8. This is consistent with the findings of \citet{Ferreras:09} who find that the number density does not evolve in the highest stellar mass bins. Below $\sim 10^{11} M_\odot$ however, our sample is likely to be incomplete due to the redshift dependant selection criteria involved in isolating LRGs. 

\begin{figure}
\begin{center}
\includegraphics[width=8.5cm,angle=0]{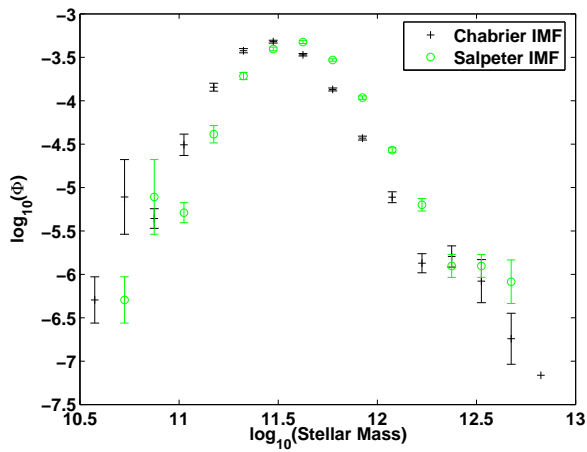}
\caption{The 2SLAQ mass function calculated using the V$_{max}$ method over the redshift range $0.4 \leq z <0.8$ when using a Salpeter and a Chabrier IMF.}
\label{fig:mf_imf}
\end{center}
\end{figure}

The discrepancy between the 2SLAQ mass function presented in this paper and the COMBO-17 mass function of \citet{Borch:06} for stellar masses greater than $\sim 10^{11} M_\odot$ may arise due to the differing choices of IMF used in the two papers. While this study uses the \citet{Chabrier:03} IMF, \citet{Borch:06} use a truncated version of the \citet{Salpeter:IMF} IMF to calculate stellar masses. In order to estimate how this would affect the mass function estimate, in Figure \ref{fig:mf_imf} we plot the 2SLAQ mass function over the entire redshift range of $0.4 \leq z < 0.8$ calculated using both the Salpeter and Chabrier IMF. The Salpeter IMF pushes the mass function towards higher masses and this results in a bigger discrepancy between the COMBO-17 and 2SLAQ data points. It is well known that the simple power-law form of the Salpeter IMF overestimates the number of low-mass stars in galaxies and therefore the stellar mass \citep{Cappellari:06, Ferreras:08} and more recent analytical forms of the IMF such as those of \citet{Chabrier:03} have fewer low-mass stars and therefore predict lower stellar masses. 

The increased number density of 2SLAQ galaxies seen at $3 \times 10^{11} M_\odot$ to $\sim 10^{12} M_\odot$ where it overlaps with COMBO-17 could be due to various reasons. Firstly, the COMBO-17 area is much smaller than that of 2SLAQ and so it is possible that there could be some incompleteness at the high-mass end of the COMBO-17 red sequence mass function due to the effects of cosmic variance. More likely however, the discrepancy probably arises due to the complicated selection of LRGs which makes it difficult to obtain a proper estimate of the volume probed. As the LRGs are selected through colour cuts in addition to the flux limit and the colour selection cuts the volume at low redshifts, the 2SLAQ galaxies will on average have a smaller $V_{max}$ than a flux-limited sample such as COMBO-17. This will result in a larger inferred number density at any given mass. 

\section{DISCUSSION}

\label{sec:disc}

In this paper we have presented luminosity and stellar mass function estimates for a sample of 8625 Luminous Red Galaxies with spectroscopic redshifts between 0.4 and 0.8 in the 2SLAQ survey. The evolution of the optical luminosity function between redshift 0.4 and 0.8 is found to be consistent with the assumptions of a passively evolving model with very little evidence for recent star formation. The number density of LRGs shows a downturn at faint magnitudes due to the effects of downsizing in the star formation whereby the less massive galaxies which are also less luminous, are bluer and therefore excluded from an LRG sample by the imposed colour selection criteria which select the reddest objects. There is also evidence that some bluer galaxies are scattered into the LRG sample due to the non-negligible photometric errors on them. 

\begin{figure}
\begin{center}
\includegraphics[width=8.5cm,angle=0]{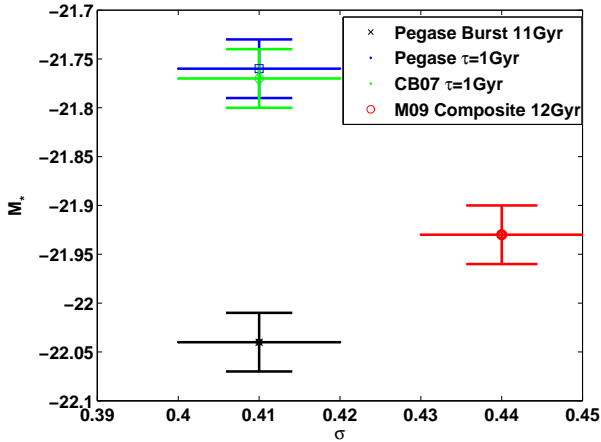}
\caption{Summary of best-fit gaussian parameters used to fit luminosity functions after assuming different spectral evolution models for k+e corrections.}
\label{fig:summary}
\end{center}
\end{figure}

The sensitivity of the optical luminosity function to changes in the spectral evolution models has also been studied. The results of this analysis are summarised in Figure \ref{fig:summary} where we show the best-fit gaussian parameters that are fit to the $V_{max}$ data points for LRGs after assuming different star formation histories in the spectral evolution models. $M_*$ moves towards brighter absolute magnitudes if we assume a Pegase burst model instead of the Maraston model, and towards fainter magnitudes if we assume a model with some residual star formation signified by a star formation timescale of $\tau=1$Gyr. The gaussians all have very similar widths although the Maraston model results in a gaussian that is slightly wider than that found with the rest of the spectral evolution models.   

The K-band luminosity function is calculated for about three quarters of the LRGs for which infra-red data is available from the UKIDSS LAS DR5. The K-band luminosity function also shows little evidence for evolution in this redshift range. The K-band luminosities are used to derive mass-to-light ratios for the LRGs by fitting CB07 spectral evolution models to the multi-band photometry. These are then used to calculate mass functions in redshift bins. Using the best-fit CB07 model instead of the Maraston models in the estimate of the mass function, makes very little difference to the mass function. We find that the most massive galaxies with $M> 3 \times 10^{11} M_\odot$ are already well assembled at redshifts of 0.8 and their number density does not change much in the redshift range considered in this work. 

\begin{figure}
\begin{center}
\includegraphics[width=8.5cm,angle=0]{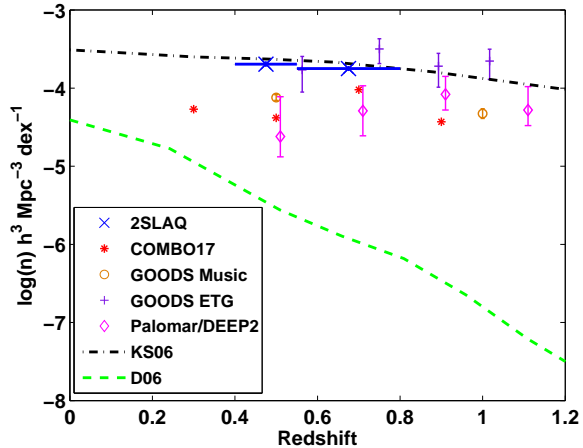}
\caption{The comoving number density of massive galaxies as a function of redshift for the 2SLAQ survey - found by integrating the mass functions presented in Figure \ref{fig:mf_imf} between $3 \times 10^{11}M_\odot$ and $10^{12}M_\odot$ along with the comoving number density from other surveys of massive galaxies as well as models of galaxy formation. KS06 refers to the \citet{Khochfar:06} model and D06 refers to the \citet{DeLucia:06} model. }
\label{fig:ndens}
\end{center}
\end{figure}

The 2SLAQ sample probes the most massive end of the mass function and can therefore be used to place stringent constraints on models of massive galaxy formation and evolution. In Figure \ref{fig:ndens} we plot the comoving number density of 2SLAQ galaxies as a function of redshift obtained by integrating the mass functions presented in Figure \ref{fig:mf}. Note that the horizontal errorbars simply represent the size of the redshift bin in our plots. We choose only galaxies with $M > 3 \times 10^{11} M_\odot$ for which we are confident that the redshift dependant selection criteria do not impose any incompleteness into the sample. We compare our estimates of the comoving number density with those from other surveys of massive galaxies as well as predictions from models of galaxy formation. All data points presented in this figure use stellar masses representative of a Chabrier IMF. Where appropriate, the quoted stellar masses in the literature have been corrected to this choice of IMF before plotting in Figure \ref{fig:ndens}. In Table \ref{tab:surveys} we summarise the samples plotted in Figure \ref{fig:ndens}. We can see that the selection criteria for all these samples is varied. In the \citet{Ferreras:09} sample for example, there is no segregation by colour but rather a visual classification of early type galaxies. In this sample, no blue galaxies are found above a stellar mass of $10^{11}M_\odot$. \citet{Pozzetti:09} use different galaxy classification schemes to derive the galaxy stellar mass function and its evolution by galaxy type and find that the massive end (M$>10^{10.5} M_\odot$) is dominated by red spheroidal galaxies upto z$\sim$1. We can therefore infer that there should be a negligible fraction of blue massive galaxies at the masses being considered in this work. We can also immediately see from Table \ref{tab:surveys}, the unique position of 2SLAQ among these surveys due to its massive volume. This allows us to significantly reduce the size of the vertical errorbars in Figure \ref{fig:ndens}.  

\begin{table*}
  \begin{center}	
    \begin{tabular}{c|c|c|c|c}
     & Redshift Range & Area & Number of Galaxies & Selection \\
          \hline
      \hline
     2SLAQ$^1$ & 0.4 - 0.8 & 186.2 deg$^2$ & 8625 & Luminous Red Galaxies \\
     COMBO-17$^2$ & 0.2 - 1.0 & 1.0 deg$^2$& $\sim$25,000 & All Galaxies \\
     Palomar/DEEP2$^3$ & 0.4 - 2.0 & 1.36 deg$^2$& 4571 & K-band selected massive galaxies \\
     GOODS-MUSIC$^4$ & 0.0 - 4.0 & 143.2 arcmin$^2$ & 2931 & K-band selected red sequence\\
     GOODS Massive Galaxies$^5$ & 0.4 - 1.2 & 320 arcmin$^2$ & 457 & Visually classified ETGs\\
      \hline
\end{tabular}	\vspace{2mm}
  \end{center}
  \caption{Summary of Massive Galaxy Samples for which stellar mass functions have been calculated - $^1$This Work, $^2$\citet{Borch:06}, $^3$\citet{Conselice:07}, $^4$\citet{Fontana:06}, $^5$\citet{Ferreras:09}. \label{tab:surveys}}
\end{table*}

From Figure \ref{fig:ndens} we can see that our work agrees well with that of \citet{Borch:06, Conselice:07} and \citet{Ferreras:09} and shows that the most massive galaxies were already well assembled at redshift 0.8 and there is little evidence for their number density having changed since then. \citet{Fontana:06} on the other hand find some evolution in the number density of massive galaxies with redshift albeit mild up to $z \simeq 1.5$. The differing comoving number densities for the different samples can be attributed to the fact that these samples have all been selected in very different ways as summarised in Table \ref{tab:surveys}. However, it is encouraging that the comoving number density for our colour selected sample of LRGs agrees well with that inferred from the visually classified sample of \citet{Ferreras:09}. 

Our observational results seem to match well predictions from the semi-analytical models of \citet{Khochfar:06}. These models follow the merging history of dark matter halos generated by the extended Press-Schechter formalism. The baryonic physics is modelled according to the prescriptions of \citet{Khochfar:05} and references therein. This model predicts that the number density of massive galaxies is almost constant upto redshifts of $\sim$1 as the number density of galaxies entering a certain mass bin from lower mass bins is counteracted by the number density of galaxies leaving that mass bin for a higher mass bin. If this were the case, we would expect the number density of galaxies in the highest mass bin to decrease as a function of redshift even if it is constant in the intermediate mass bins. We compute the comoving number density in the higher mass bin of $12 < \log_{10}(M_*/M_\odot) < 12.5$ and in our two redshift bins - $0.4 \leq z < 0.55$ and $0.55 \leq z < 0.8$. The number densities per unit dex in stellar mass thus obtained are $\log(n)/h^3 Mpc^{-3} dex^{-1}=-5.39 \pm ^{0.10} _{0.13}$ and  $\log(n)/h^3 Mpc^{-3} dex^{-1}=-5.55 \pm ^{0.08} _{0.09}$ respectively. Therefore there is no evidence that the comoving number density of the most massive galaxies is decreasing with increasing redshift. The agreement between our data points and the \citet{Khochfar:06} models shown in Figure \ref{fig:ndens} should therefore be treated with caution as the constant number density of massive galaxies in the models, is only maintained due to a fine-tuning of objects between mass bins and we find no evidence that this fine-tuning actually occurs.  

Our results also do not match at all those generated from models based on the Millennium simulations \citep{DeLucia:06}. In these models, AGN feedback is invoked to shut off star formation after a characteristic mass scale in order to reproduce the observed colour-bimodality and luminosity function at redshift 0. As a consequence, the downsizing in star formation is reproduced but the growth of massive galaxies at high redshifts of greater than 0.5 is prohibited and dry mergers become the most dominant mechanism for mass growth in such galaxies. This is clearly inconsistent with our observations which suggest not only that significant star formation in massive galaxies has ceased at a redshift of 0.8 but also that these galaxies were already well assembled at these redshifts. The main difference between the \citet{DeLucia:06} and \citet{Khochfar:06} models is the feedback with respect to AGN activity rather than the fact that the former follows a numerical simulation for the evolution of the dark matter halos whereas the latter relies on an analytical approach. If cooling is suppressed, the efficiency of massive galaxy formation at early epochs is much lower accounting for the lower number densities in the \citet{DeLucia:06} models. We also note that the \citet{DeLucia:06} models are computed from the Millennium simulations, which use a cosmology with $\sigma_8$=0.9, whereas the \citet{Khochfar:06} models have been updated for this work with the later WMAP5 cosmology with $\sigma_8$=0.8 \citep{Dunkley:WMAP5}. As the AGN activity affects the star formation history of the galaxy, we conclude that the star formation and quenching mechanisms invoked in the different galaxy formation models need to be revisited in light of our new constraints on the very high mass end of the stellar mass function. 

\citet{Granato:04} on the other hand have proposed a way in which to bring about the anti-hierarchical formation of the baryonic component of galaxies in models while still working within the framework of the $\Lambda$CDM cosmology. This is essentially done by invoking feedback from supernovae as well as the nuclear activity in massive galaxies. These processes slow down star formation in the least massive halos and drive gas outflows thereby increasing the stellar to dark matter ratio in the more massive halos and ensuring that the physical processes acting on baryons are able to effectively reverse the order of formation of galaxies compared to dark matter halos. These models have already been shown to match the local K-band luminosity function of massive galaxies \citep{Granato:04} as well as the luminosity function at redshifts of $\sim$1.5 \citep{Granato:04, Silva:05}. Reproducing our observational results for the most massive galaxies at intermediate redshifts would therefore be an important test for these models.  

Recently \citet{Dekel:09} have also proposed an alternative galaxy formation model whereby bursts of star formation can occur in massive galaxies without the need for violent merging events. These authors suggest that the massive galaxies, which reside in the centers of filaments, are fed by cold gas streams that penetrate the shock heated media of massive dark matter halos thereby inducing star formation. 

Given the small errorbars in our estimates of the comoving number density compared to previous studies and the fact that our sample has secure spectroscopic redshifts for all objects and a massive area that reduces errors due to cosmic variance, this new observational result presents a significant challenge for many current models of galaxy formation and requires a reinvestigation of the feedback mechanisms involved. 

\section{CONCLUSION}

This paper has examined the evolution of the optical and near infra-red luminosity function as well as the stellar mass function of 8625 LRGs from the 2SLAQ survey between redshift 0.4 and 0.8. We have demonstrated the unique position of 2SLAQ LRGs among other massive galaxy samples due to the large volume probed by the sample as well the availability of spectroscopic redshifts for all galaxies and near infra-red data for $\sim$75\% of them. This has allowed us to probe the very high-mass end of the stellar mass function which places the most stringent constraints on models of galaxy formation. We have also used the spectral evolution models of \citet{Maraston:09} which are the most accurate models of LRGs to date to study the evolution of these objects although our conclusions change little if other models are used instead. Specifically we draw the following conclusions: 

\begin{itemize}

\item{The evolution of the optical luminosity function of LRGs between redshifts 0.4 and 0.8 is consistent with the assumptions of a passive evolution model where the stars were formed at high redshift with little or no evidence for recent episodes of star formation.}

\item{A downturn is seen at the faint end of the LRG luminosity function due to the effects of downsizing in the star formation. The faintest galaxies are expected to be less massive and therefore bluer than the bright objects and such objects are excluded from the LRG sample due to the imposed colour selection of these objects which selects only the reddest galaxies.}

\item{We find that photometric errors may induce a significant bias into our sample of LRGs and scatter galaxies both in and out of our sample across the colour selection boundaries. We show that more galaxies are likely to be scattered into the sample than are scattered out leading to an overprediction of the observed space density but the shape of the luminosity function does not change if we remove the galaxies with large photometric errors.} 

\item{The LRG luminosity function is found to be unaffected by cosmic variance due to the large volume occupied by the sample.}

\item{The luminosity function is also relatively insensitive to the choice of spectral evolution model and the metallicity although models with some residual star formation shift the luminosity function towards fainter absolute magnitudes.}

\item{The stellar mass function for these LRGs for $M > 3 \times 10^{11} M_\odot$ also shows little evidence for evolution between redshifts 0.4 and 0.8 suggesting that these most massive systems were already well assembled at redshifts of 0.8. This is consistent with the emerging picture of downsizing in the mass assembly of massive galaxies.}

\item{The stellar mass function estimate is also relatively insensitive to the choice of spectral evolution model assumed in the calculation of $V_{max}$. However, different choices of the stellar initial mass function will shift the mass function estimate as found in previous studies.}

\item{The comoving number density of LRGs with $M> 3 \times 10^{11} M_\odot$ has changed little between redshifts 0.8 and 0.4.  The same is true for LRGs with $M > 10^{12} M_\odot$. This is consistent with other observational results for massive galaxy samples and does not agree with the predictions of most current galaxy formation models. We find that the models of \citet{Granato:04} may be promising in matching our observations although they are yet to be compared with massive galaxy samples at intermediate redshifts.} 

\end{itemize}

Overall, our results support the emerging picture of downsizing in both the star formation and mass assembly of early type galaxies and present a significant challenge for current models of galaxy formation. We have shown our findings to be robust to changes in spectral evolution models, cosmic variance and photometric errors and conclude that these new observations of the most massive and most luminous galaxies in our Universe will need to be reconciled with the models if progress is to be made in the field of massive galaxy formation and evolution. 

\section*{Acknowledgements}

We thank David Wake for many useful discussions. We would also like to thank Eduardo Cypriano and Samuel Farrens for productive discussions at the early stages of this work and Daniel Mortlock and Mat Page for their help with the UKIDSS data. MB acknowledges support from the Science and Technology Facilities Council (STFC). FBA acknowledges support from the Leverhulme Trust via an Early Careers Fellowship. 


\bibliography{}

\begin{appendix}

\section{The STY Estimator}

The most common parametric estimator of the luminosity function is that of \citet{STY:79} (STY). A parametric form is assumed for the luminosity function e.g. a gaussian as in this paper. The parameters, $\bf{a}$ are then solved for by maximising the likelihood: 

\begin{equation}
\mathcal{L}(\bf{a})=\prod_{i} p_i
\label{eq:likelihood}
\end{equation}

\noindent where $p_i$ is the probability of a galaxy with absolute magnitude, $M$ being selected in the flux limited, colour cut sample and is given by Eq \ref{eq:pi}

\begin{equation}
p_i=\frac{\Phi(M_i,\bf{a})f(M_i,z_i)}{\int \Phi(M, \bf{a}) f(M_i, z_i) dM}
\label{eq:pi}
\end{equation}

\noindent where $f(M_i,z_i)$ is the apparent magnitude incompleteness function for objects with absolute magnitude, $M_i$ and redshift, $z_i$. This incompleteness occurs due to various reasons. Firstly, not all galaxies targetted in a survey will have redshifts determined succesfully. \citet{Wake:06} have considered this incompleteness function for 2SLAQ LRGs using carefully constructed 2SLAQ survey masks. Although they find a slight dependance of this incompleteness function on the (g-r) colour, including this functional dependance as opposed to considering a constant incompleteness of 76.1\% introduces less than a 1\% change in their luminosity function estimate. We therefore assume a constant redshift incompleteness of 76.1\% that is independant of both the redshift and the absolute magnitude. 

\begin{figure}
\begin{center}
\includegraphics[width=8.5cm,angle=0]{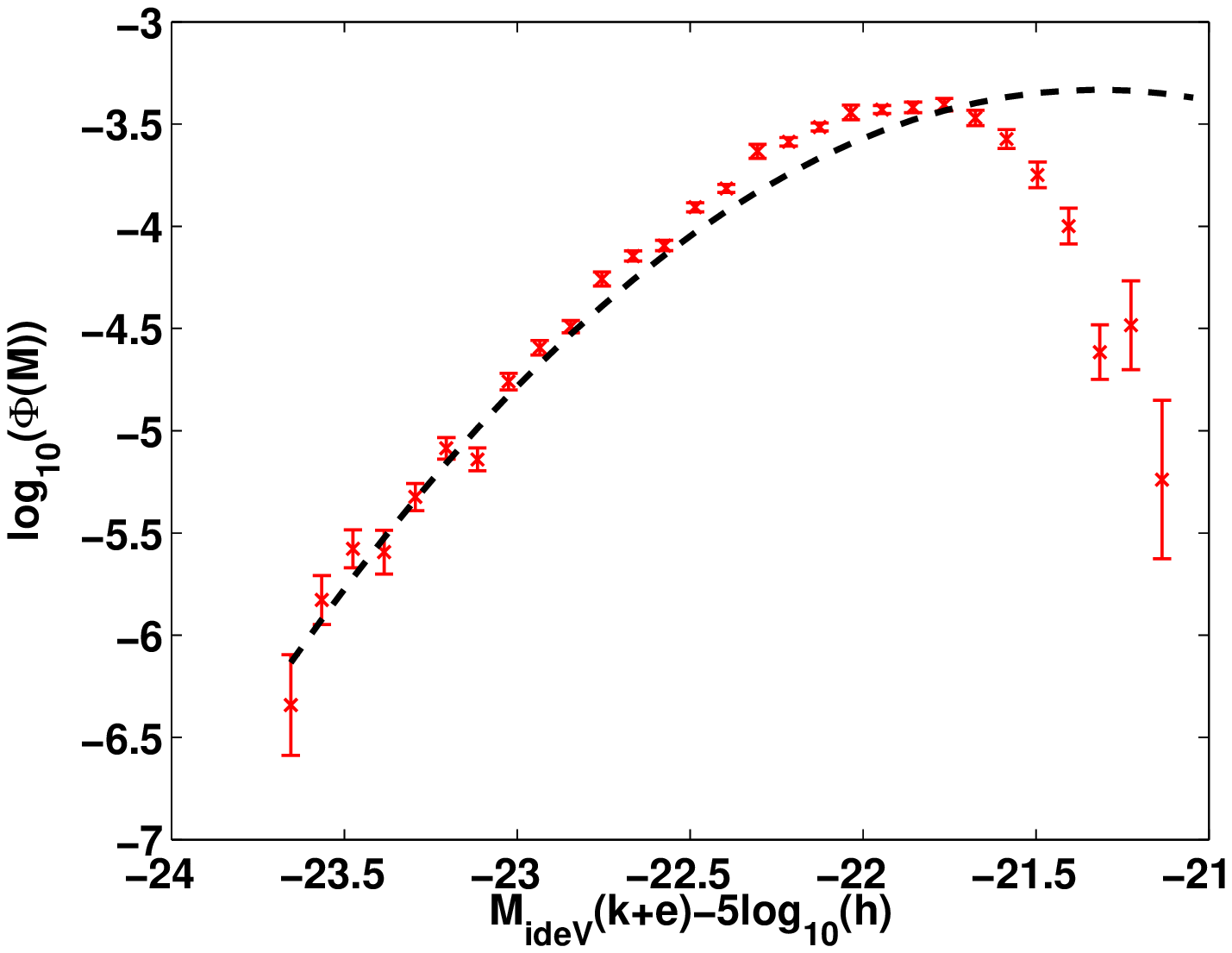}
\caption{The non-parametric $V_{max}$ data points for the LRGs in the entire sample along with the STY gaussian fit to the entire sample.}
\label{fig:sty}
\end{center}
\end{figure} 

In the case of a colour-selected sample such as 2SLAQ however, there may be additional sources of incompleteness as the space density of these galaxies is a function of both the flux limit and the colours. The 2SLAQ LRGs at different redshifts occupy different distributions in the colour absolute magnitude plane due to the redshift dependant colour selection. This colour selection could introduce an incompleteness function that is both a function of absolute magnitude as well as redshift and this selection function has to be explicitly included in the integral in Eq \ref{eq:pi}. We have checked that using the absolute magnitude averaged selection function rather than the 2d function $f(M_i,z_i)$, makes very little difference to the STY estimate of the luminosity function. In this case, the galaxy incompleteness can be taken outside the integral and the likelihood in the STY method adjusted directly e.g. \citep{Zucca:94, Willmer:06}. 

\begin{equation}
\mathcal{L}(\bf{a})=\prod_{i} p_i^{w_i}
\label{eq:likelihood2}
\end{equation}

\noindent where the weights associated with each individual galaxy are $w_i = 1/S(m_{i})$. We initially calculate luminosity functions assuming $w_i=1$ for the entire sample and this is shown by the dashed line in Figure \ref{fig:sty}. This clearly illustrates that the V$_{max}$ and STY estimators differ considerably from each other at the faint-end. The STY fit parameters are $M_*=-21.32, \sigma=0.65, \phi_*=9.97\times10^{-4}$ for the entire sample and these values are very different to those derived from the $\chi^2$-fits in $\S$ \ref{sec:lf}. It has already been noted in this paper that fainter than $i\sim19.3$, photometric errors may become significant in scattering LRGs both in and out of the colour selection boundaries. This analysis therefore suggests that the STY estimator is particularly sensitive to photometric errors for colour selected samples and the weights $w_i$ start to differ significantly from 1 for galaxies with large photometric errors. In order to calculate these weights, one would have to do a Monte Carlo simulation of galaxies assuming a certain spectral evolution model, add noise to this simulation and then look at the completeness function $S(m_{i})$ for each galaxy by considering which galaxies are selected on applying the 2SLAQ colour selection criteria e.g. \citep{Fried:01, Wolf:03}. This is the subject of future work. The STY estimator in its traditional form is clearly biased for a colour-selected sample with non-negligible photometric errors and for this reason, it has not been presented in the main body of this paper. We note that when we move to high redshift bins where the galaxies are generally brighter and the photometric errors smaller, the STY estimator and $V_{max}$ estimator agree very well as it is clearly reasonable in these regimes to assume $w_i=1$. 

\end{appendix}

\end{document}